\documentclass[prd,twocolumn,showpacs,preprintnumbers,amsmath]{revtex4}[11pt]
\usepackage{graphicx}	
\usepackage{amssymb}	
\usepackage{amsmath}	
\usepackage{mathptm}	
\usepackage{multirow}
\usepackage{array}
\usepackage{hyperref}
\usepackage{mathrsfs}
\usepackage{color}	

\newcommand{\be}{\begin{equation}}
\newcommand{\ee}{\end{equation}}
\newcommand{\bea}{\begin{eqnarray}}
\newcommand{\eea}{\end{eqnarray}}

\begin{document}



\title{Excited States of U(1)$_{2+1}$ Lattice Gauge Theory from Monte Carlo Hamiltonian}

\author{
A.~Hosseinizadeh$^{a}$\footnote{
Corresponding author, Email:
ahmad.hosseinizadeh.1@ulaval.ca},
G.~Melkonyan$^{a}$,
H.~Kr\"{o}ger$^{a,b}$,
M.~McBreen$^{a,c}$,
N.~Scheu$^{a}$
}

\affiliation{
$^{a}$ {\small\sl Physics Department, Laval University, Qu\'{e}bec (QC) G1V 0A6, Canada} \\
$^{b}$ {\small\sl Frankfurt Institute for Advanced Studies, Goethe Universit\"at Frankfurt, 60438 Frankfurt am Main, Germany\footnote{present address}} 
\\
$^{c}$ {\small\sl Department of Mathematics, Fine Hall, Princeton University, Princeton, NJ 08544-1000, USA\footnote{present address}} 
}



\begin{abstract}

We address an old problem in lattice gauge theory - the computation of the spectrum and wave functions of excited states. Our method is based on the Hamiltonian formulation of lattice gauge theory.
As strategy, we propose to construct a stochastic basis of Bargmann link states, drawn from a physical probability density distribution. Then we compute transition amplitudes between stochastic basis states. From a matrix of transition elements we extract energy spectra and wave functions. We apply this method to U(1)$_{2+1}$ lattice gauge theory. We test the method by computing the energy spectrum, wave functions and thermodynamical functions of the electric Hamiltonian of this theory and compare them with analytical results. We observe a reasonable scaling of energies and wave functions in the variable of time. We also present first results on a small lattice for the full Hamiltonian including the magnetic term.

\end{abstract}


\maketitle



%
\section{Introduction}
\label{sec:Intro}
%
Since the invention of lattice gauge theory (LGT) by Wilson in 1974~\cite{Wilson74}, it has been customary to compute Euclidean 2-point functions (and likewise n-point functions), which is the expectation value of two field operators in the vacuum-to-vacuum transition amplitude,
\be
\label{eq:NPointFunction}
\langle \Omega, t=+\infty | \Psi(x) \Psi(y) | \Omega, t=-\infty \rangle ~ .
\ee
The term Euclidean means that one works in imaginary time $t \to -it$. It is important to note that the ground state wave function $\Omega$ is actually not known in LGT. The ground state wave function emerges in the above amplitude via the Feynman-Kac limit through projection from an arbitrary state $|\Xi \rangle$ onto the vacuum state in the limit of large Euclidean time,
\be
\label{eq:FeynmanKac}
\exp[-\mathcal H\;t/\hbar] ~ | \Xi \rangle \sim \lim_{t \to \infty}
\exp[-E_{gr} t/\hbar] | \Omega \rangle \langle \Omega | \Xi \rangle ~ .
\ee
If one wants to construct the wave functions of the ground state or excited states, one needs  information beyond the vacuum-to-vacuum amplitude. Thus let us consider transition amplitudes
\be
\label{eq:TransAmplUpsilon}
\langle \Upsilon_{\nu}, t=T | \Upsilon_{\mu}, t=0 \rangle =
\langle \Upsilon_{\nu} | \exp[-\mathcal H T/\hbar ] | \Upsilon_{\mu} \rangle
\ee
between initial and final states taken from some set of states $|\Upsilon_{\nu} \rangle, \nu=1,\dots,N$. Here $|\Upsilon_{\nu} \rangle$ denotes time-independent Bargmann-link states, that is, a configuration of link variables $U$ assigned to each link $ij$ on the whole {\it spatial} lattice. As a reminder, in the case of a chain of coupled oscillators, a Bargmann state denotes the ensemble of displacements of the particles from their resting positions. Hence Bargmann states can be interpreted as the analogue of position states in quantum mechanics.

It is crucial to choose the states $|\Upsilon_{\nu} \rangle$ to be physically relevant and important. In the history of quantum mechanics it has been a tradition to use basis states obeying mathematical properties, like being orthogonal functions and forming a complete basis. Examples are
Fourier functions, Hermite functions, etc. These functions have no relation to the particular physical system, e.g. a hydrogen atom. This is a luxury which one can afford when solving a system with few degrees of freedom. However, in many-body physics and quantum field theory, with a huge (infinite) number of degrees of freedom, one better chooses a basis suited to the particular physical system.

There is a long tradition in many-body nuclear physics, condensed matter physics and elementary particle physics aiming to compute wave functions and the energy spectrum from matrix elements of a Hamilton operator in a suitable basis. This approach has been successful in special cases, e.g.,
when perturbation theory is applicable (nuclear shell model, Kondo effect~\cite{Wilson75}), or in situations where few degrees of freedom describe physics (Schwinger model in 1+1 dimensions, renormalization group Hamiltonian for critical phenomena). However, in general, this approach resulted in mitigated success. The problem is due to the choice of basis states. This can be understood by the example of diagonalization of a real symmetric matrix of large but finite rank, which is ill-conditioned (large difference in order of magnitude between largest and smallest eigenvalue). Any set of orthogonal basis states will yield a few leading order eigenvalues. However, only basis states close to the eigen basis will resolve eigenvalues beyond leading order.

Carrying over this lesson to many-body physics and field theory, it is not surprising that perturbatively constructed basis states (e.g. Fock states) in general are not adequate, i.e., those states do not reflect the physically important degrees of freedom. This goes hand in hand with another potentially serious problem: Is there a controllable cut-off? The Tamm-Dancoff cut-off gives no estimate of the remainder. Also, it can happen (e.g., in the coupled cluster method) that one has a quite large set of basis states (in the order of $10^3$ to $10^4$) yielding many tiny matrix elements all of about the same order of magnitude. Where to draw the line?

We suggest a solution to such problems lying in a combination of two strategies: (i) Use stochastic techniques to sample states from a huge variety of possibilities. (ii) Use information from the physical system to guide the sampling. The above principles (i,ii) mean to do Monte Carlo with importance sampling. This procedure has proven most successful in LGT to compute path integrals via generation of equilibrium path configurations. The stochastic basis states will be closely related to those equilibrium path configurations.

From computation of transition amplitudes between stochastic basis states, one obtains a spectrum and wave functions of an effective Hamiltonian - the
so-called Monte Carlo Hamiltonian - being valid in a low energy, respectively low-temperature window. The idea of the Monte Carlo Hamiltonian has been
suggested in 1999 in Ref.~\cite{Jirari99}. Its working in quantum mechanics has been demonstrated in a number of cases. For the 1-D harmonic oscillator,
energy spectrum, wave functions and thermodynamical functions have been found to be in good agreement with the exact results~\cite{Jirari99,Jirari00,Kroger00}. Similar results have been obtained for uncoupled as well as coupled harmonic oscillators in 2-D~\cite{Jiang00,Luo00a} and 3-D~\cite{Luo01a}. This has been extended to a variety of other potentials in 1-D like $V \propto x^{2} + x^{4}$~\cite{Luo00b}, $V \propto |x|/2$, and $V \propto \theta(x) x$~\cite{Luo02}, as well as the $1/r$ Coulomb potential with a singularity at the origin~\cite{Luo04}. The Monte Carlo Hamiltonian has been applied to the Yukawa potential $V=-V_{0} ~ exp(-\alpha r)/r$ in the search for a critical value of $\alpha_{c}$ above which
no bound states exist~\cite{Li06}. In field theory, the Monte Carlo Hamiltonian has been applied to the $(1+1)$ Klein-Gordon model for the computation of the spectrum and thermodynamical functions~\cite{Caron01,Luo01b,Luo01c,Kroger03b}, and likewise to the $(1+1)$ scalar model for the computation of the spectrum and thermodynamical functions~\cite{Huang02,Kroger03a,Kroger07}. A first step towards the Monte Carlo Hamiltonian in lattice gauge theory has been made in~\cite{Paradis07} by computing transition amplitudes of U(1) gauge theory.

In lattice gauge theory the U(1) model in $(2+1)$ dimensions has been investigated since the early days using Euclidean Monte Carlo methods. Bhanot and
Creutz~\cite{Bhanot80}, D'Hoker~\cite{Hoker81}, Ambjorn et al.~\cite{Ambjorn82} computed the Wilson loop. Sterling and Greensite~\cite{Sterling83} computed the string tension from energy differences using external sources. The Hamiltonian formulation of the U(1)$_{2+1}$ model has been investigated using a variety of different methods. We can refer to the Quantum Monte Carlo methods~\cite{Chin84,Yung86,Hamer94,Hamer00},
the projector Monte Carlo method~\cite{Potvin84,Dahl85}, the ensemble projector Monte Carlo method~\cite{DeGrand85}, the Green's function Monte Carlo method~\cite{Heys83}, the Langevin technique~\cite{Eleuterio87} and the guided random walk method~\cite{Kotchan90}. Furthermore, the model has been studied via the t-expansion method~\cite{Horn87,Morningstar92}, the block renormalization group method~\cite{Lana88}, the correlated basis function method~\cite{Dabringhaus91}, the coupled-cluster expansion~\cite{Fang96,Baker96}, and series expansions~\cite{Hamer92,Hamer96}. The U(1)$_{2+1}$ model has been also investigated in the finite lattice Hamiltonian approach by Allesandrini et al.~\cite{Alessandrini82}, and Irving et al.~\cite{Irving83}.
Recent results from the path integral Monte Carlo approach have been reported in~\cite{Loan03}.

In this work we construct the Monte Carlo Hamiltonian in U(1)$_{2+1}$ lattice gauge theory and apply it to compute the energy spectrum of excited states, the corresponding wave functions and thermodynamical functions. The goal of this article is to show the working of the method applied to an Abelian, but non-trivial model in lattice gauge theory. We consider a few small spatial lattice volumes up to $8^{2}$. Because of the smallness of the lattices we do not consider here the quantum continuum limit ($a \to 0$). Rather, we try to give a careful analysis of the origin and size of errors, which determine the limitations of the method. We believe the potential of the method lies in its capacity to determine energies and wave functions for a number of excited states.

%
\section{Monte Carlo Hamiltonian in quantum mechanics}
%
%
\begin{figure}[h,t]
\centering
\includegraphics[width=60mm, height=75mm]{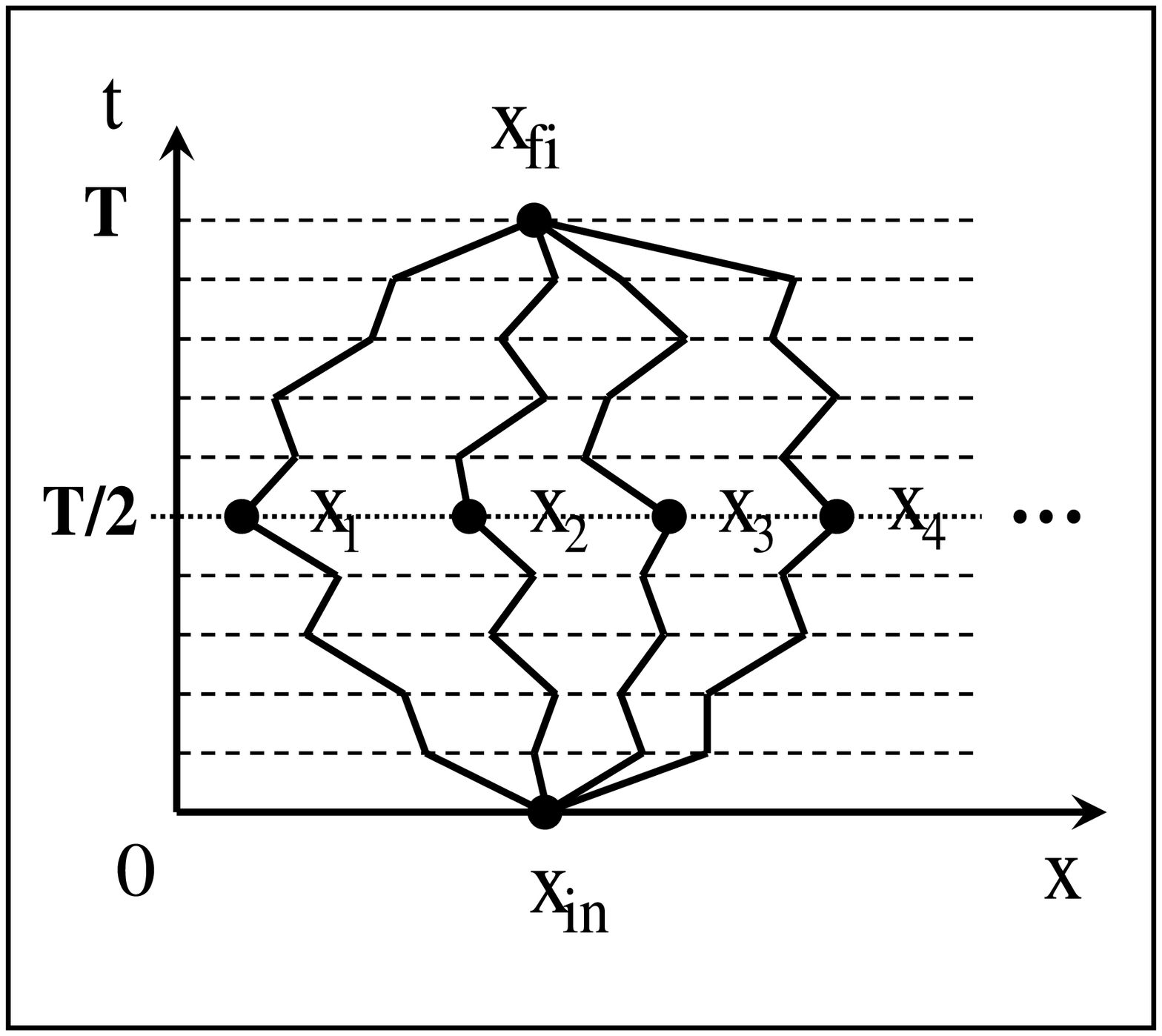}\\
\includegraphics[width=85mm, height=45mm]{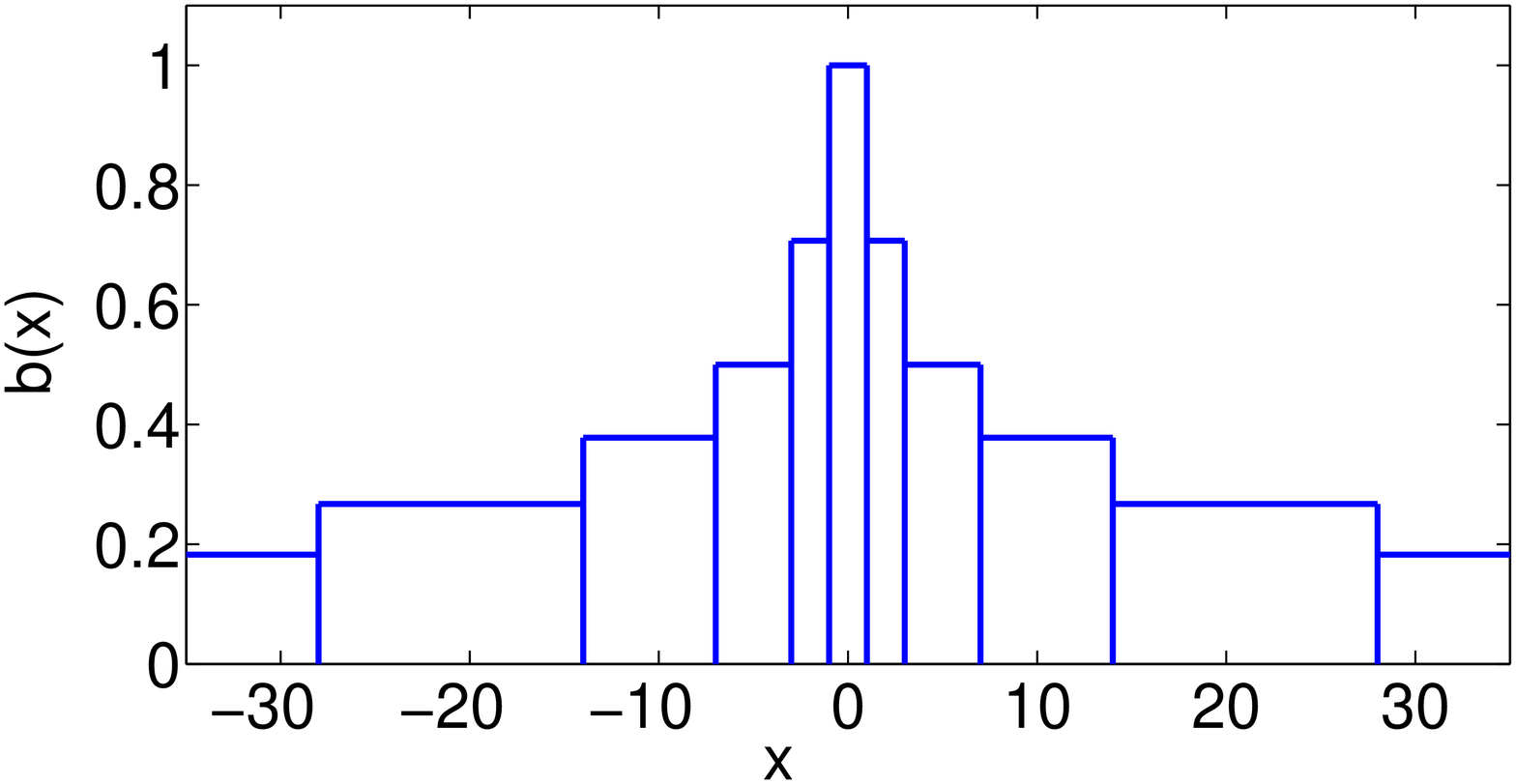}
\caption{Quantum mechanics. Top panel: typical paths from path integral. Bottom panel: Scheme of box functions. The position of the center of any box function is drawn from a Gaussian distribution. They form a stochastic basis.}
\label{fig:QMpath+boxfct}
\end{figure}
Let us consider in 1-D a massive particle moving in a potential $V(x)$. Let $S$ denote its classical action and let $\mathcal H$ denote its quantum mechanical Hamilton operator. First we construct a stochastic basis, adapted to the particular physical system. We do this by a random draw of nodes $x_{i}$ guided by the action $S$, respectively, Hamiltonian operator $\mathcal H$ of the system. We draw from the probability distribution, given by the transition amplitude in imaginary time
\be
\label{eq:ProbDistr}
P(x) = \mathcal Z ~ \langle x | \exp[-\mathcal H T/\hbar ] | x=x_\text{init} \rangle ~ ,
\ee
where $\mathcal Z$ is a normalization factor such that $\int dx ~ P(x) = 1$. The function $P(x)$ is positive, because the transition amplitude, expressed as path integral is given by a positive measure $[dx]$ and a positive weight factor $\exp[-S_\text{Eucl}/\hbar]$ ($S_\text{Eucl}$ denotes the Euclidean action). For example, in case of the free Hamiltonian, choosing $x_\text{init}=0$, the distribution is a Gaussian,
\begin{eqnarray}
\label{eq:FreeProbDistr}
P(x) &=& \langle x | \exp[-\mathcal H_\text{kin} T/\hbar ] | x_\text{init}=0 \rangle
\nonumber \\
&=& \sqrt{\frac{m}{2 \pi \hbar T } }
\exp[ - \frac{1}{\hbar} \frac{m}{2} x^{2}/T] ~ .
\end{eqnarray}
Here, $\mathcal H_\text{kin}$ commutes with generators of translation and the measure $dx$ is translationally invariant. In case of the harmonic oscillator, $P(x)$ is also given by a Gaussian. In order to draw samples from the distribution $P(x)$ given by Eq.~(\ref{eq:ProbDistr}), one can use the path integral in imaginary time. For example, Fig.~\ref{fig:QMpath+boxfct}~(top panel) schematically shows the sampling of points $x_{i}$ from the distribution $P(x)$ with the time parameter $t=T$, via the path integral going from $t=0$ to $t=T$, by intersecting paths at $t=T/2$. In such a way one obtains a set of nodes $\{ x_{1}, \dots, x_{N}\}$. This yields a stochastic basis of (non-normalizable) position states $|x_{1} \rangle, \dots, |x_{N} \rangle $. In order to eventually compute normalized wave functions of the Hamiltonian, one needs a basis of normalized Hilbert states. We introduce a stochastic basis of square integrable Hilbert states ("box states"), defined by
\be
b_{i}(x)= \left\{
\begin{array}{ll}
h_{i} & \mbox{if} ~~~ x_{i} -\Delta x_{i}/2 < x <
x_{i} +\Delta x_{i}/2
\\
0 & \mbox{else}
\end{array}
\right.
~ .
\ee
Those states are "quasi-localized", i.e., the function $b_{i}(x)$ is located around the node $x_{i}$ and has a "small" width $\Delta x_{i}$.
The normalization condition of the distribution, $\int dx ~ P(x) = 1$, implies
\be
\sum_{i=1}^{N} \Delta x_{i} ~ P(x_{i}) \approx 1 ~ .
\ee
Thus we define
\be
\Delta x_{i} = \frac{1}{N P(x_{i})} ~ .
\ee
The normalization condition of the box functions, $\int dx ~ b_{i}^{2}(x) = 1$, yields
\be
h_{i}^{2} \Delta x_{i} = 1 ~ .
\ee
Thus one obtains
\be
h_{i} = \frac{1}{\sqrt{\Delta x_{i}}} = \sqrt{N P(x_{i})} ~ .
\ee
Fig.~\ref{fig:QMpath+boxfct} schematically shows the stochastic basis of box states.

Next, let us consider the computation of transition amplitudes in imaginary time between a pair of initial and final position states $|y \rangle$ and $|z \rangle$, respectively.
\bea
\label{eq:TransAmplQM}
M_{z,y} = \langle z | \exp[-\mathcal H T/\hbar] | y \rangle
\nonumber \\
\left. = \int [dx] \exp[- S[x]/\hbar ] \right|^{z,T}_{y,0} ~ .
\eea
Here $S[x]$ denotes the Euclidean action of a path $x(t)$ going from $y$ to $z$.
\be
S[x] = \left. \int_{0}^{T} dt ~ \frac{1}{2} m \dot{x}^{2} + V(x) \right|_{y}^{z} .
\ee
The Monte Carlo method with importance sampling is suited and conventionally applied to estimate a ratio of path integrals. Thus we suggest to calculate the matrix elements $M_{z,y}$ by splitting the action
\be
S = S_{0} + S_{V} \equiv
\int_{0}^{T} dt ~ \frac{1}{2} m \dot{x}^{2} +
\int_{0}^{T} dt ~ V(x) ,
\ee
and to express $M_{z,y}$ as
\bea
\label{eq:TransAmplPathInt}
&& M_{z,y}(T)
\left. = \int [dx] ~ \exp[ -S_{0}[x]/\hbar ] \right|^{z,T}_{y,0}
\times
\nonumber \\
&&\times \frac{
\left. \int [dx] ~ \exp[ - S_{V}[x]/\hbar ] ~ \exp[ -S_{0}[x]/\hbar ] \right|^{z,T}_{y,0}
}
{
\left. \int [dx] ~ \exp[ -S_{0}[x]/\hbar ] \right|^{z,T}_{y,0}
}
\nonumber \\
&& \equiv M^{(0)}_{z,y}(T) \times R_{z,y}(T) ~ .
\eea
Here $M^{(0)}_{z,y}$ denotes the matrix elements corresponding to the free action $S_{0}$, which is known analytically, that is, 
\be
M^{(0)}_{z,y}(T) =
\sqrt{ \frac{ m }{ 2 \pi \hbar T } } ~
\exp \left[ - \frac{ m}{2 \hbar T } (z - y)^{2} \right] .
\ee
The term $R_{z,y}(T)$ denotes the ratio of path integrals. Such ratio can be computed by standard Monte Carlo methods with importance sampling, by treating $\hat O \equiv \exp[ -S_{V}/\hbar]$ as observable.

Next, we want to compute transition amplitudes in imaginary time between a pair of initial and final states taken from the stochastic basis of box states. Thus we build the finite, real, symmetric matrix
\bea
\label{eq:TransAmplBoxStates}
M_{ji}(T)
&=& \langle b_{j} | \exp[-\mathcal H T/\hbar] | b_{i} \rangle
\nonumber \\
&=& h_{j} h_{i}
\int_{x_{j}-\Delta x_{j}/2 }^{x_{j}+\Delta x_{j}/2} d z
\int_{x_{i}-\Delta x_{i}/2}^{x_{i}+\Delta x_{i}/2} d y ~
M_{z,y}(T) ~ ,
\nonumber \\
\eea
where $i,j$ run over $1,\dots,N$. Using the property that box states are quasi-local, and assuming the transition amplitudes vary little in the domain covered by the box functions, we may approximate $M_{ij}$ in the following way
\bea
\label{eq:ApproxTransAmplBoxStates}
M_{ji}(T) &\approx&
h_{j} h_{i} ~ \Delta x_{j} \Delta x_{i} ~
M^{(0)}_{x_{j},x_{i}}(T) \times R_{x_{j},x_{i}}(T)
\nonumber \\
&=& \sqrt{\Delta x_{j} \Delta x_{i} } ~
M^{(0)}_{x_{j},x_{i}}(T) \times R_{x_{j},x_{i}}(T) ~ .
\nonumber \\
\eea
After having obtained the matrix $M(T)$, given by Eq.~(\ref{eq:ApproxTransAmplBoxStates}), one can extract eigenvalues and wave functions, which define an effective Hamiltonian. $M(T)$ is a positive and Hermitian matrix, under the assumption that $\mathcal H$ is a Hermitian operator. Elementary linear algebra implies that there is a unitary matrix $U$ and a real, diagonal matrix $\mathcal D$ such that
\be
\label{eq:DiagonalMatrix}
M(T) = U^{\dagger} ~ \mathcal D(T) ~ U .
\ee
On the other hand, projecting $\mathcal H$ onto the subspace $S_{N}$ generated by the first $N$ states of the basis $|b_{i} \rangle$, and using the eigen representation of such Hamiltonian, one has
\be
\label{eq:SpectralRepMatrix}
M_{ji}(T) =
\sum_{k=1}^{N} \langle b_{j} | E^\text{eff}_{k}  \rangle e^{-E^\text{eff}_{k} T/\hbar}  \langle E^\text{eff}_{k} | b_{i}  \rangle ,
\ee
and we can identify
\be
\label{eq:DefEnergyWaveFct}
U^{\dagger}_{ik} =  \langle b_{i} | E^\text{eff}_{k} \rangle, ~~~ \mathcal D_{k}(T) = e^{-E^\text{eff}_{k} T/\hbar } .
\ee
Thus algebraic diagonalization of the matrix $M(T)$ yields eigenvalues $\mathcal D_{k}(T), ~ k=1, \cdots, N$, which by Eq.~(\ref{eq:DefEnergyWaveFct})
determines the spectrum of energies,
\be
\label{eq:EnergEigValue}
E^\text{eff}_{k} = - \frac{\hbar}{T} \ln \mathcal D_{k}(T), ~~ k= 1, \cdots, N .
\ee
The corresponding k-th eigenvector can be identified with the k-th column of the matrix $U^{\dagger}_{ik}$. Hence from Eq.~(\ref{eq:DefEnergyWaveFct}) we also obtain the wave function of the k-th eigenstate expressed in terms of the basis $| b_{i} \rangle$. Thus starting from the matrix elements $M_{ij}(T)$ we have explicitly constructed an effective Hamiltonian in a diagonal form
\be
\mathcal H^\text{eff} = \sum_{k =1}^{N} | E^\text{eff}_{k}  \rangle E^\text{eff}_{k} \langle E^\text{eff}_{k} | ~ .
\ee
%

%
\section{U(1)$_{2+1}$ lattice gauge theory}
%
The strategy to construct an effective Hamiltonian, outlined above for 1-D quantum mechanics, can be translated to lattice gauge theory. However, there are some major differences, which one encounters in doing so. First and foremost, U(1)$_{2+1}$ gauge theory is based on the principle of local gauge symmetry, absent in 1-D quantum mechanics. As a consequence, physical states (wave functions of physical particles) have to be gauge invariant states. The lattice action $S[U]$ is given by~\cite{Loan03}
\begin{eqnarray}
\label{eq:SplitGaugeAction}
S[U] &=& \frac{1}{g^{2}}\frac{a}{a_{0}} \sum_{\Box_{t}}
[1 -Re(U_{\Box})]
\nonumber \\
&+& \frac{1}{g^{2}}\frac{a_{0}}{a} \sum_{\Box_{s}}
[1 -Re(U_{\Box})]
\nonumber \\
&\equiv&  S_\text{elec}[U] + S_\text{mag}[U] ~ ,
\end{eqnarray}
where $\Box_{t}$ and $\Box_{s}$ denote the time-like and space-like plaquettes, respectively. The corresponding lattice Hamiltonian is given by~\cite{Irving83}
\be
\label{eq:GaugeHamilton}
\mathcal H = \frac{g^{2}}{2a} \sum_{<ij>} \hat{l}_{ij}^{2} + \frac{1}{g^{2}a} \sum_{\Box_{s}} [1 - Re(U_{\Box})] ~ .
\ee
The lattice Hamiltonian has two terms, the electric term and a magnetic term,
\bea
\mathcal H_\text{elec} &=& \frac{g^{2}}{2a} \sum_{<ij>} \hat{l}_{ij}^{2}
\nonumber \\
\mathcal H_\text{mag} &=& \frac{1}{g^{2} a} \sum_{\Box_{s}} [1-Re(\hat{U}_{\Box})]  ~ .
\eea
The operator $\hat{l}_{ij}$ counts the number of electric flux strings. Its eigenstates are
\begin{equation}
\label{eq:ElectField}
\hat{l}_{ij} |\lambda_{ij} \rangle = \lambda_{ij} |\lambda_{ij} \rangle ~ , \lambda_{ij} = 0,\pm 1, \pm 2, \dots ~ .
\end{equation}
For each link $ij$, the states $|\lambda \rangle$ form a complete orthogonal basis,
\bea
&&\sum_{\lambda = 0,\pm1,\dots} | \lambda \rangle \langle \lambda | = 1 ~ ,
\nonumber \\
&& \langle \lambda' | \lambda \rangle = \delta_{\lambda',\lambda} ~ .
\eea
The magnetic term is built from link operators $\hat{U}_{ij}$. It has the eigenstates
\be
\hat{U}_{ij} | U_{ij} \rangle = {U}_{ij} | U_{ij} \rangle ~ .
\ee
For each link $ij$, also this basis is a complete orthogonal basis,
\bea
\int dU ~ |U \rangle \langle U | = 1 ~ ,
\nonumber \\
\langle U' | U \rangle = \delta(U'-U) ~ .
\eea
The quantum mechanical analogue of the electric flux string is momentum $P$, while the analogue of the link is the position $X$. Sometimes one needs to switch from the Bargmann link basis to the electric field string basis. In quantum mechanics the scalar product $\langle x|p \rangle$ can be computed from the commutator $[\hat{X},\hat{P}] =i ~ \hbar$. Likewise, the scalar product $\langle \lambda | U \rangle$ can be obtained from the commutator~\cite{Paradis07}
\be
[\hat{l}, \hat{U}] = - \hat{U} ~ .
\ee
One obtains
\begin{equation}
\label{eq:ScalarProd}
 \langle\lambda|U \rangle = (U)^{\lambda} ~ .
\end{equation}

Secondly, the relation between the transition amplitude at one hand expressed in terms of the Hamiltonian and at the other hand expressed in terms of the
path integral (in analogy to Eq.~(\ref{eq:TransAmplQM})), translates in gauge theory to
\bea
\label{eq:TransAmplGauge}
M_{U_\text{fi},U_\text{in}} = \langle U_\text{fi} | \exp[-\mathcal H T/\hbar] | U_\text{in} \rangle
\nonumber \\
\left. = \int [dU] \exp[- S[U]/\hbar ] \right|^{U_\text{fi},T}_{U_\text{in},0} ~ .
\eea
The Hamiltonian requires to choose a gauge and the Hamiltonian has been obtained using the temporal gauge ($U_{time-like}=1$). However, Eq.~(\ref{eq:TransAmplGauge}) is not true, in general. This is due to the fact that a Bargmann link state $| U \rangle $ is not a gauge invariant state and hence, the amplitude expressed in terms of the Hamiltonian is not gauge invariant either. However, the amplitude expressed in terms of the path integral is gauge invariant, because the Haar measure $dU$ is invariant and the action $S[U]$ is also gauge invariant. Thus, both amplitudes can
{\it not} be equal, in general. However, one can show that they become equal after projection of the Bargmann link states onto gauge invariant states~\cite{Scheu}, that is
\bea
\label{eq:TransAmplGaugeInv}
\langle U_\text{fi} | \hat{\Pi} \exp[-\mathcal H T] | U_\text{in} \rangle
\left. = \int [dU] \exp[- S[U] ] \right|^{U_\text{fi},T}_{U_\text{in},0} ~ .
\nonumber \\
\eea
Here, the operator $\hat{\Pi}$ denotes a projection operator of Bargmann states onto gauge invariant Bargmann states, which commutes with the Hamiltonian. For a single link $\hat{\Pi}$ is given by
\be
\label{eq:ProjectorGaugeInv}
\hat{\Pi} | U_{ij} \rangle = \int d\mathscr G_{i} d\mathscr G_{j} ~ | \mathscr G_{i} U_{ij} \mathscr G^{-1}_{j} \rangle ~ ,
\ee
and correspondingly for a multiple link state ($\mathscr G_{i}$ and $\mathscr G_{j}$ are group elements).

Thirdly, we would like to point out that in lattice gauge theory, it is customary to identify the measure $[dU]$ in Eq.~(\ref{eq:TransAmplGauge}) with the Haar measure of the group. This is not quite right, as there is some normalization factor missing. Such normalization factor, however, cancels out when computing ratios of path integrals as is usually done in computing observables via Monte Carlo with importance sampling. However, if the path integral stands alone, such normalization factor needs to be taken into account. In the case of U(1) lattice gauge theory, such normalization factor has been computed in Ref.~\cite{Paradis05}. It is given as follows. We consider the Euclidean transition amplitude for a single time slice $a_{0}$ on the lattice under the electric part of the Hamiltonian (like in quantum mechanics such normalization factor only depends on the kinetic term). The integral of the transition amplitude over all final states (which is gauge invariant) is normalized to unity,
\begin{eqnarray}
\label{eq:LGTAmplProbInterpret}
&&\int dU_\text{fi} \langle U_\text{fi}|\exp[- \mathcal H_\text{elec} a_{0} ]| U_\text{in} \rangle
\nonumber \\
&=& \int [\mathcal Z\;dU_\text{fi}] \exp[-S_\text{elec}[U]|_{U_\text{in},t=0}^{U_\text{fi},t=a_{0}}
\nonumber \\
&=& \int_{- \pi}^{+\pi}\mathcal Z ~ \frac{d \alpha_\text{fi} }{2\pi}
\exp \left[- \frac{a}{g^{2} a_{0}}
[1 - \cos(\alpha_\text{fi} - \alpha_\text{in})] \right]
= 1 ~ .
\nonumber \\
\end{eqnarray}
Defining $A= a/(g^{2}a_{0})$, and using the Bessel function of imaginary argument~\cite{Gradshteyn},
\begin{equation}
I_{0}(z) = \frac{1}{\pi} \int_{0}^{\pi} d\theta \exp[z\cos(\theta)] ~ ,
\end{equation}
Eq.~(\ref{eq:LGTAmplProbInterpret}) yields
\begin{equation}
\label{eq:ZGeneral}
\mathcal Z(A) = \frac{\exp(A)}{I_{0}(A)} ~ .
\end{equation}
We keep $T$ fixed, and let ${N\to\infty}$ (which means $a_{0}\to 0$ and $A\to \infty$), which is the continuum limit in time direction. However, in space direction we keep $a=1$. The asymptotic behavior of $\mathcal Z$ for $a_{0}$ going to zero is given by
\begin{eqnarray}
\label{eq:ZFactor}
\mathcal Z(A) = \sqrt{2 \pi A} [1 -\frac{1}{8} A^{-1} - \frac{7}{128} A^{-2}
+ O(A^{-3})] ~ .
\nonumber \\
\end{eqnarray}
In the limit $a_{0}\to 0$ the leading term $\mathcal Z(A) = \sqrt{2 \pi A}$ is sufficient to guarantee that the amplitude under the integral in
Eq.~(\ref{eq:LGTAmplProbInterpret}) goes over to $\delta(U_\text{fi} - U_\text{in})$, as should be. A numerical simulation~\cite{Paradis05} has shown that in lattice gauge theory (with non-zero $a_{0}$) also the sub-leading term is important and can not be neglected.

%
\section{Stochastic basis}
%
We want to find a finite set of physically relevant basis states $\{\Upsilon_{\mu} \mid \mu=1,\dots,N]\}$ in Bargmann space. We suggest to choose those states randomly, drawn from a physically guided distribution. There are several possibilities in choosing such distribution. For a system with a given (Kogut-Susskind) Hamiltonian $\mathcal H$ the physically motivated choice for the distribution is given by the transition amplitude in imaginary time, involving the same Hamiltonian,
\be
\label{eq:TrialProbDistr}
P(U) = \langle U | \exp[\mathcal H T/\hbar ] | U_\text{in} \rangle ~ ,
\ee
where $U_\text{in}$ is some suitably chosen fixed spatial lattice configuration (Bargmann state). This function is suitable as probability distribution because it is a positive function $P(U) \ge 0$. The positivity can be seen by expressing the transition amplitude in Eq.~(\ref{eq:TrialProbDistr}) in terms of an Euclidean path integral, where the group measure is positive and the exponential $\exp[ -S[U]/\hbar ]$ is also positive. This choice of distribution has the inconvenience of not being analytically computable. Nevertheless, physically relevant configurations (Bargmann states) can be drawn from this distribution by expression $P(U)$ as path integral like in quantum mechanics (see above) and doing the sampling via Monte Carlo method.
Still considering the underlying Hamiltonian $\mathcal H$, one may consider as alternative the distribution given by the transition amplitude from the electric Hamiltonian,
\be
\label{eq:DefProbDistr}
P(U) = \langle U | \exp[-\mathcal H_\text{elec} T/\hbar ] | U_\text{in} \rangle ~ .
\ee
This distribution analytically is a computable function. Moreover, this function is normalized to unity, $\int dU P(U) = 1$.  The latter property is due to the fact that first the Hamiltonian $\mathcal H_\text{elec}$ is a Casimir operator for gauge transformation (commutes with all generators of local gauge transformations) and second due to the left/right invariance of the group measure. The analytical computation of the distribution $P(U)$ given by Eq.~(\ref{eq:DefProbDistr}) is discussed below. Using Monte Carlo technique to sample from such distribution $P(U)$, one obtains the Bargmann states denoted by $\mid \Upsilon_{\mu} \rangle$. Both of the above alternatives of distribution involve a time parameter $T$, which determines the "width" of the distribution. Such time parameter needs to be tuned. As a general rule, we used to choose $T$ such that it falls into the scaling window of eigenvalues (see below).

%
\section{Transition amplitudes under electric Hamiltonian}
\label{sec:AnalExpr}
%
Let us consider the transition amplitude between Bargmann link states (for simplicity we consider first a lattice consisting of a single spatial link) under evolution of the electric Hamiltonian, $\langle U_\text{fi}|\exp[- \mathcal H_\text{elec} T/\hbar ]| U_\text{in} \rangle$. Its quantum mechanical analogue is $\langle x_\text{fi}|\exp[- \mathcal H_\text{kin} T/\hbar ]| x_\text{in} \rangle $. In elementary quantum mechanics one learns to compute the latter amplitude by expanding position states in terms of momentum states which are eigen states of the kinetic Hamiltonian. This is done via the Fourier expansion theorem. One can proceed by analogy to compute the above amplitude in lattice gauge theory. The position $x$ corresponds to the link variable
$U$. Because $U=\exp[i \alpha]$ is a periodic function in the interval $-\pi \le \alpha \le \pi$, the conjugate variable of $U$ has discrete values, being the eigenvalues of electric flux strings, Eq.~(\ref{eq:ElectField}). Thus functions of the variable $U=\exp[i \alpha]$ can be expanded by discrete Fourier expansion. In case of the Abelian group U(1), the Peter-Weyl theorem says that the Bargmann link states can be expanded in terms of irreducible representation matrices. In this case, the Peter-Weyl theorem is equivalent to Fourier expansion~\cite{Tung}. The Peter-Weyl theorem holds more generally for groups SU(N)~\cite{Vilenkin} (see Appendix). Applying the Peter-Weyl theorem for the group U(1) to the above amplitude yields
\begin{eqnarray}
\label{eq:PeterWeylU1}
&&\langle U_\text{fi}|\exp[- \mathcal H_\text{elec} T/\hbar ]| U_\text{in}\rangle
\nonumber \\
&=& \sum_{n=0,\pm1,\pm2,\dots}
\exp[- \frac{g^{2} \hbar T}{2a} n^{2} ]
(U_\text{fi}^{-1} U_\text{in})^{n} ~ .
\nonumber \\
\end{eqnarray}
In mathematical terms $n$ (running over $0,\pm1,\pm2,\dots$) denotes the index of the irreducible representation. $(U)^{n}$ also denotes the irreducible representation of group element $U$ with representation index (quantum number) $n$. In physical terms, $n$ represents the number of electric flux lines (see Eq.~(\ref{eq:ElectField})). The Hamiltonian $\mathcal H_\text{elec}$ is a Casimir, which is diagonal in the representation index $n$. Note that the amplitude only depends on the group elements via the product $U_\text{fi}^{-1} U_\text{in}$. By parameterizing the link variables via $U = \exp[i \alpha]$, one obtains
\begin{eqnarray}
\label{eq:HamTimeEvol}
&&<U_\text{fi}|\exp[- \mathcal H_\text{elec} T/\hbar ]| U_\text{in}>
\nonumber \\
&=& \sum_{n=0,\pm1,\pm2,\dots}
\exp[- \frac{g^{2} \hbar T}{2a} n^{2} ]
\cos[n(\alpha_\text{fi} - \alpha_\text{in})] ~ .
\nonumber \\
\end{eqnarray}

%
\subsection{Construction of gauge invariant states}
\begin{figure}[h,t]
\centering
\includegraphics[width=0.5\linewidth,angle=0]{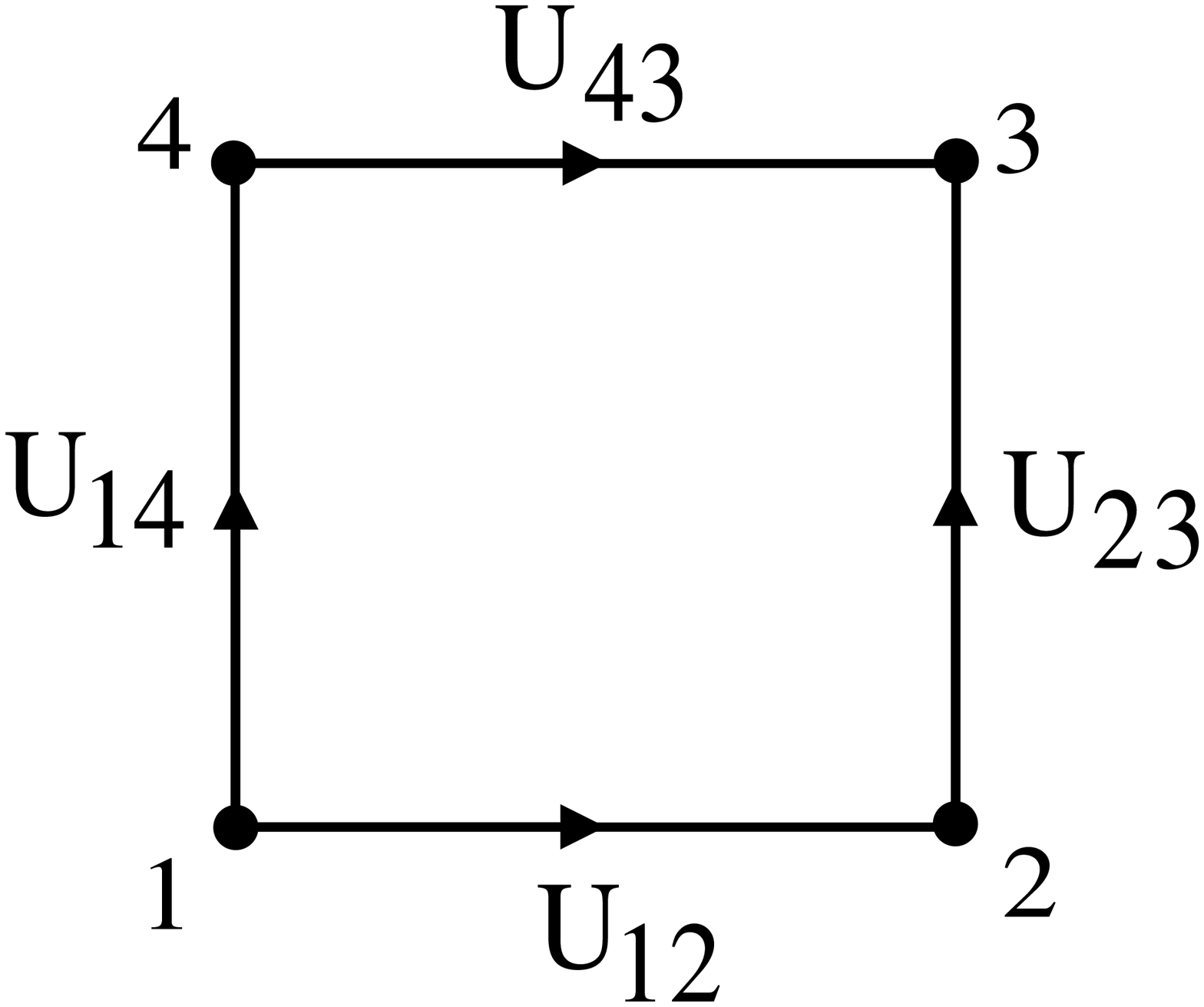} \\
\includegraphics[width=0.5\linewidth,angle=0]{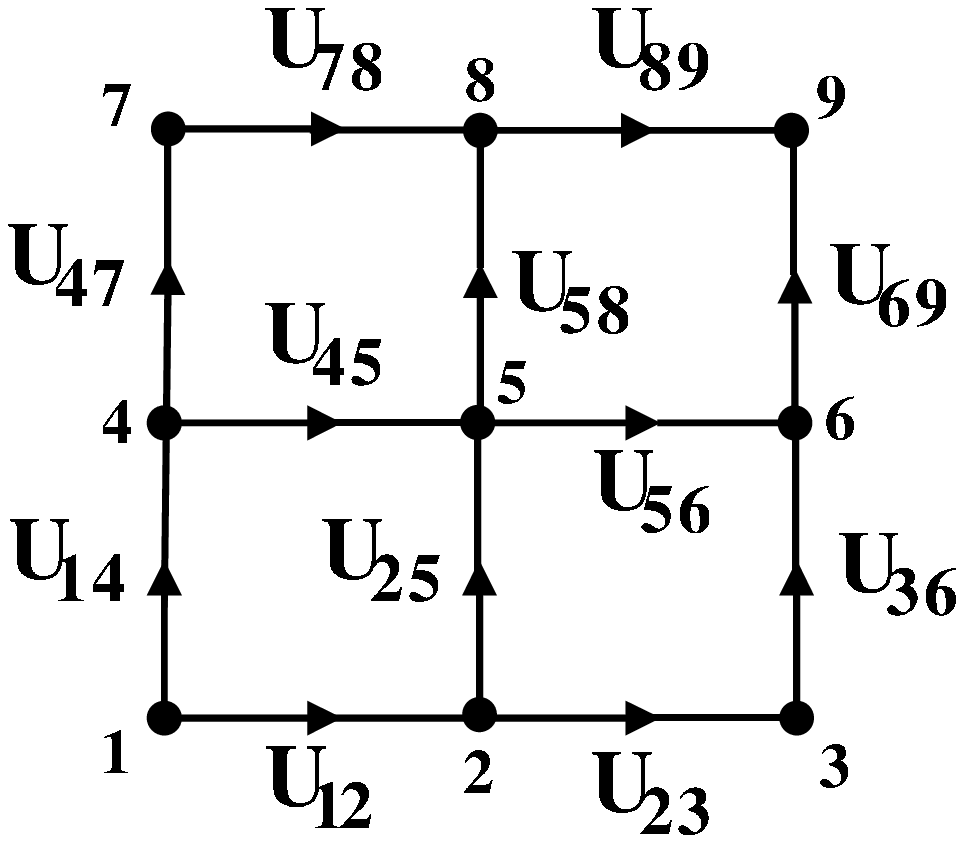} \\
\includegraphics[width=0.9\linewidth,angle=0]{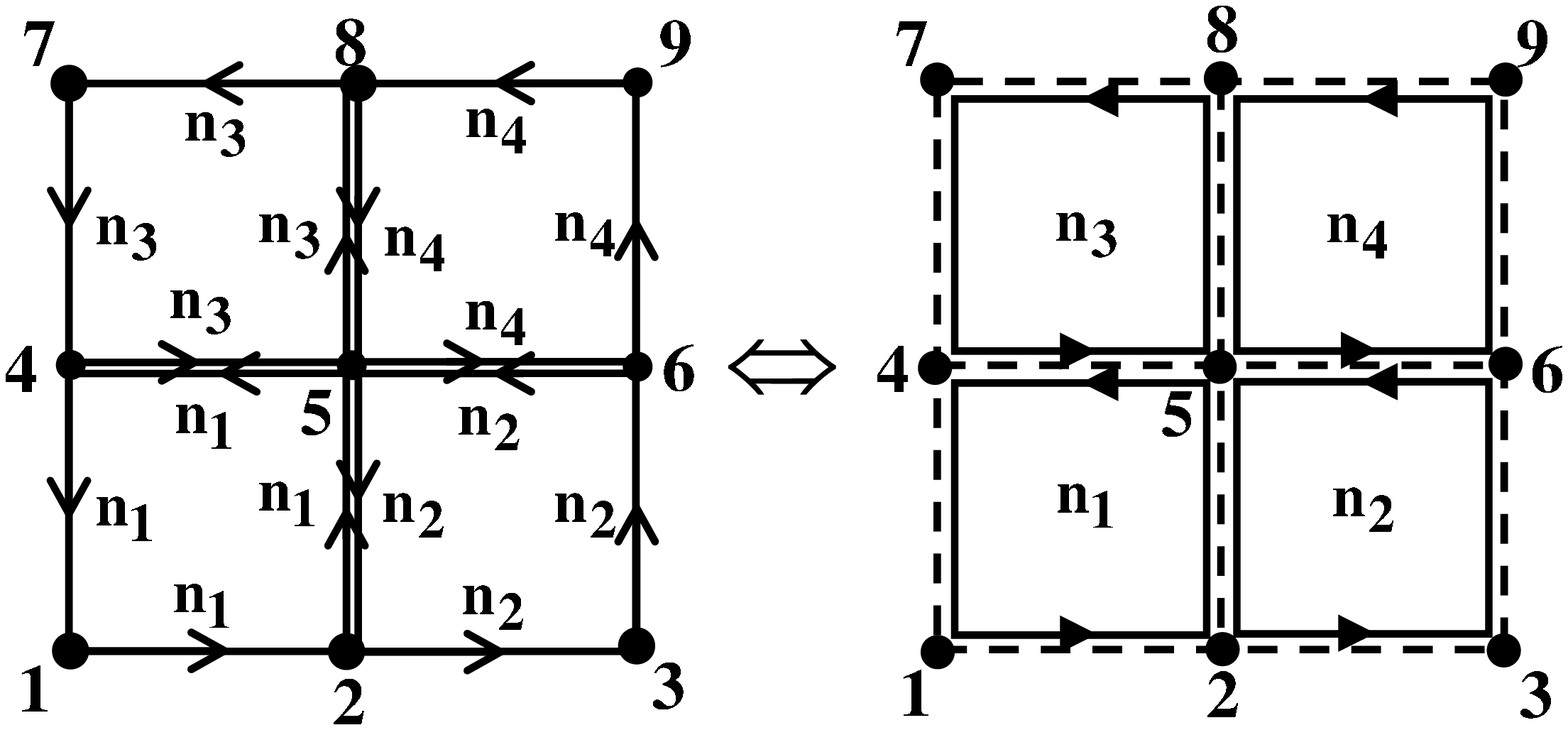} 
\caption{Spatial lattice. 
Top panel: The state of a spatial plaquette described by Bargmann link states $|U_{ij}\rangle$.
Middle panel: the same scheme for a lattice including four plaquettes. 
Bottom panel: The state of lattice can also be expressed in terms of electric flux states, by assigning a link flux number $n_{ij}$ and a direction of flux lines between neighbor sites $i$ and $j$. A gauge invariant state obeys Gauss' law at each vertex, i.e., the total number of incoming minus outgoing flux lines is zero. On a 2-D lattice, a gauge invariant state can be  characterized by assigning a plaquette flux number $n_{pl}$ plus an orientation to each
plaquette.}
\label{fig:4Plaquettes}
\end{figure}

%
The Peter-Weyl theorem can be applied to an arbitrary lattice. It is useful for the construction of gauge invariant states. For example, let us take a spatial lattice consisting of four links ordered to form a plaquette (see Fig.~\ref{fig:4Plaquettes}, top panel). Then the transition amplitude is given by
\begin{eqnarray}
\label{eq:1PlaqAmpl}
&&\langle U^\text{fi}| \exp[-\mathcal H_\text{elec}T/\hbar] | U^\text{in} \rangle
\nonumber \\
&\equiv& \langle U^\text{fi}_{12},U^\text{fi}_{23},U^\text{fi}_{43},U^\text{fi}_{14}
| e^{(-\mathcal H_\text{elec}T/\hbar)} |
U^\text{in}_{12},U^\text{in}_{23},U^\text{in}_{43},U^\text{in}_{14} \rangle
\nonumber \\
&=& \prod_{ij=12,23,43,14} ~\bigg\{ \sum_{n_{ij}=0,\pm1,\pm2,\dots} \exp \left[ - \frac{g^{2}\hbar T}{2a} n^{2}_{ij} \right]
\nonumber \\
&\times&
\cos [n_{ij}(\alpha^\text{fi}_{ij} - \alpha^\text{in}_{ij})] \bigg\} ~ .
\end{eqnarray}
In order to make the amplitude gauge invariant, we carry out the group integral $\int d \mathscr G_{i} = \frac{1}{2\pi} \int_{o}^{2\pi} d \beta_{i}$
at the nodes $i=1,2,3,4$, as
\begin{eqnarray}
\label{eq:1PlaqAmplGaugeInv}
&&\langle U^\text{fi} |\hat{\Pi} ~ \exp[-\mathcal H_\text{elec}T/\hbar] | U^\text{in} \rangle
\nonumber \\
&=&
\left(\frac{1}{2\pi}\right)^{4} \int_{o}^{2\pi} d \beta_{1}
\dots \int_{o}^{2\pi} d \beta_{4}
\nonumber \\
&\times&
\prod_{ij=12,23,43,14} ~ \sum_{n_{ij}=0,\pm1,\pm2,\dots} ~
\exp \left[ - \frac{g^{2}\hbar T}{2a} n^{2}_{ij} \right] ~
\nonumber \\
&\times& \cos [n_{ij}(\alpha^\text{fi}_{ij} - (\alpha^\text{in}_{ij} + \beta_{i} - \beta_{j}))]  ~ .
\end{eqnarray}
The group integral introduces Kronecker delta functions at each vertex. E.g., at vertex $j$, one obtains $\delta_{n_{ij},n_{jk}}$. The number of ingoing flux lines equals to the number of outgoing flux lines. As we consider here the absence of any charge, this rule represents Gauss' law.
By defining the plaquette angle
\begin{eqnarray}
&&\theta_\text{plaq} = \alpha_{12} + \alpha_{23} + \alpha_{43} + \alpha_{14} ~ ,
\nonumber \\
&&\Delta \theta_\text{plaq} = \theta^\text{fi}_\text{plaq}  - \theta^\text{in}_\text{plaq}  ~ ,
\end{eqnarray}
we obtain the final expression of the gauge invariant amplitude,
\begin{eqnarray}
\label{eq:1PlaqAmplGaugeInv2}
&&\langle U^\text{fi}_\text{inv} | \exp[-\mathcal H_\text{elec}T/\hbar] | U^\text{in}_\text{inv} \rangle
\nonumber \\
&=&
\sum_{n=0,\pm1,\pm2,\dots} \exp \left[ - \frac{g^{2}\hbar T}{2a} 4 n^{2} \right] ~
\cos \Bigl[ n \Delta \theta_\text{plaq}  \Bigr] ~ .
\nonumber \\
\end{eqnarray}
Here $n$ denotes the number of closed plaquette loops on top of each other. The result is built from plaquettes which are closed loops of consecutive link variables, forming the smallest non-local gauge invariant objects on the lattice. Note: Opposite signs of $n$ correspond to plaquettes of opposite orientation. The eigenvalue of the electric field $\vec{E}^{2}$ corresponds to the contribution from $n$ plaquette loops. It is also important to note that the result only depends on the number of plaquette loops and the difference between initial and final plaquette angles.

This result can be generalized to any 2-D spatial lattice including $N_P$ plaquettes, as following:
\begin{eqnarray} 
&&\langle U^\text{fi}_\text{inv}
| \exp[-\mathcal H_\text{elec}T/\hbar] |
U^\text{in}_\text{inv} \rangle
\nonumber \\
&=&
\prod_{P=1}^{N_{P}} \bigg\{\sum_{n_P=0,\pm1,\pm2,\dots} \exp \left[ - \frac{g^{2}\hbar T}{2a} E^{2}_\text{graph} \right]
\nonumber 
\\
&\times & 
\cos \Bigl[ \vec n_P.\vec{\Delta\theta_P} \Bigr] \bigg\}~ .
\label{eq:AnalyticFormulae}
\end{eqnarray}
For example, in the case of a spatial lattice composed of four plaquettes (Fig.~\ref{fig:4Plaquettes}, middle and bottom panels) the gauge invariant amplitude becomes
\begin{eqnarray}
\label{eq:4PlaqAmplGaugeInv}
&&\langle U^\text{fi}_\text{inv}
| \exp[-\mathcal H_\text{elec}T/\hbar] |
U^\text{in}_\text{inv} \rangle
\nonumber \\
&=&
\sum_{n_1,n_2,n_3,n_4=0,\pm1,\pm2,\dots} \exp \left[ - \frac{g^{2}\hbar T}{2a} E^{2}_\text{graph} \right]
\nonumber \\
&\times&
\cos \Bigl[ n_1\; \Delta \theta^\text{plaq}_{1254} + n_2\; \Delta \theta^\text{plaq}_{2365} +
n_3\; \Delta \theta^\text{plaq}_{5698} + n_4 \;\Delta \theta^\text{plaq}_{4587} \Bigr] ~ ,
\nonumber \\
\end{eqnarray}
where
\begin{eqnarray}
&& E^{2}_\text{graph} = 2n_1^{2} + 2n_2^{2} + 2 n_3^{2} + 2n_4^{2} +
\nonumber \\
&&+ (n_1-n_2)^{2} + (n_1-n_3)^{2} + (n_2-n_3)^{2} + (n_3-n_4)^{2} ~.
\nonumber \\
\end{eqnarray}
$E^{2}_\text{graph}$ denotes the eigenvalue of the electric field operator of the graph constructed from plaquettes filling the lattice, each
with counter-clockwise orientation.
Analytic results have been obtained also in three-dimensional spatial lattices, where instead of plaquettes one has cubic states. Because we present numerical results in 2-D lattices, we do not give details of 3-D here.
%
%
%
%
\begin{table}[h,t]
\caption{Electric Hamiltonian. Comparison of eigenvalues of effective Hamiltonian
versus exact eigenvalues. Lattice $2^2$, spacing $a=a_0=1$,
time parameter in distribution $P(U)$ is $T=0.1$, coupling $g=1$,
dimension of stochastic basis $N_\text{basis}=32$.}
\label{tab:ElecHamiltoni}
\centering
\begin{tabular}{|l|c|l|c|l|}
\hline
$n$ & $\mathcal D_{n}$  & ${E_n}^\text{(eff)}$  & ${E_n}^\text{(exact)}$ & Rel. error \\
\hline \hline
 0 & ~1.0000000000~~ & ~~0.0000000 & ~~0.0 & ~~~~~~---  \\
 1 & 0.8187308431 & ~~1.9999989 & ~~2.0 & 5.5$\times$ 10$^{-7}$ \\
 2 & 0.8187306875 & ~~2.0000008 & ~~2.0 & 4.0$\times$ 10$^{-7}$ \\
 3 & 0.4493290180 & ~~7.9999988 & ~~8.0 & 1.5$\times$ 10$^{-7}$ \\
 4 & 0.4493289101 & ~~8.0000012 & ~~8.0 & 1.5$\times$ 10$^{-7}$ \\
 5 & 0.1652988898 & ~17.9999999 & ~18.0 & 5.5$\times$ 10$^{-9}$ \\
 6 & 0.1652988667 & ~18.0000013 & ~18.0 & 7.2$\times$ 10$^{-8}$ \\
 7 & 0.0407622060 & ~31.9999995 & ~32.0 & 1.5$\times$ 10$^{-8}$ \\
 8 & 0.0407621941 & ~32.0000024 & ~32.0 & 7.5$\times$ 10$^{-8}$ \\
 9 & 0.0067379553 & ~49.9999876 & ~50.0 & 2.5$\times$ 10$^{-7}$ \\
 10 & 0.0067379496 & ~49.9999961 & ~50.0 & 7.8$\times$ 10$^{-8}$ \\
 11 & 0.0007466034 & ~71.9997633 & ~72.0 & 3.3$\times$ 10$^{-6}$ \\
 12 & 0.0007466023 & ~71.9997789 & ~72.0 & 3.1$\times$ 10$^{-6}$ \\
 13 & 0.0000554676 & ~97.9971144 & ~98.0 & 2.9$\times$ 10$^{-5}$ \\
 14 & 0.0000554663 & ~97.9973409 & ~98.0 & 2.7$\times$ 10$^{-5}$ \\
 15 & 0.0000027819 & 127.9235530 & 128.0 & 6.0$\times$ 10$^{-4}$ \\
 16 & 0.0000027792 & 127.9332884 & 128.0 & 5.2$\times$ 10$^{-4}$ \\
 17 & 0.0000001401 & 157.8084005 & 162.0 & 2.6$\times$ 10$^{-2}$ \\
 18 & 0.0000001134 & 159.9218144 & 162.0 & 1.3$\times$ 10$^{-2}$ \\
 19 & 0.0000000873 & 162.5307945 & 200.0 & 1.9$\times$ 10$^{-1}$ \\
 20 & 0.0000000810 & 163.2828884 & 200.0 & 1.8$\times$ 10$^{-1}$ \\
\hline
\end{tabular}
\end{table}
%
%
\section{Numerical results}
\label{sec:NumResults}
%
\subsection{Test of effective electric Hamiltonian}
%
In order to test the effective Hamiltonian, first we consider the electric Hamiltonian. This is a good test bed, because the physics of the electric Hamiltonian can be computed analytically. First and as an example, for the case of a $2^{2}$ lattice, we present the eigenvalues $\mathcal D_{n}$ of the matrix $M$ and the energy eigenvalues $E_{n}$ of the effective Hamiltonian versus the exact eigenvalues, as well as the relative errors (see Tab.~\ref{tab:ElecHamiltoni}). The following behavior is observed. The ground state, corresponding to $n_\text{plaq}=0$ (zero plaquette loops), has no degeneracy. All excited states are doubly degenerated, corresponding to $n_\text{plaq}=\pm 1, \pm2, \dots$. 
 \begin{figure}
 \centering
 \includegraphics[width=85mm, height=40mm]{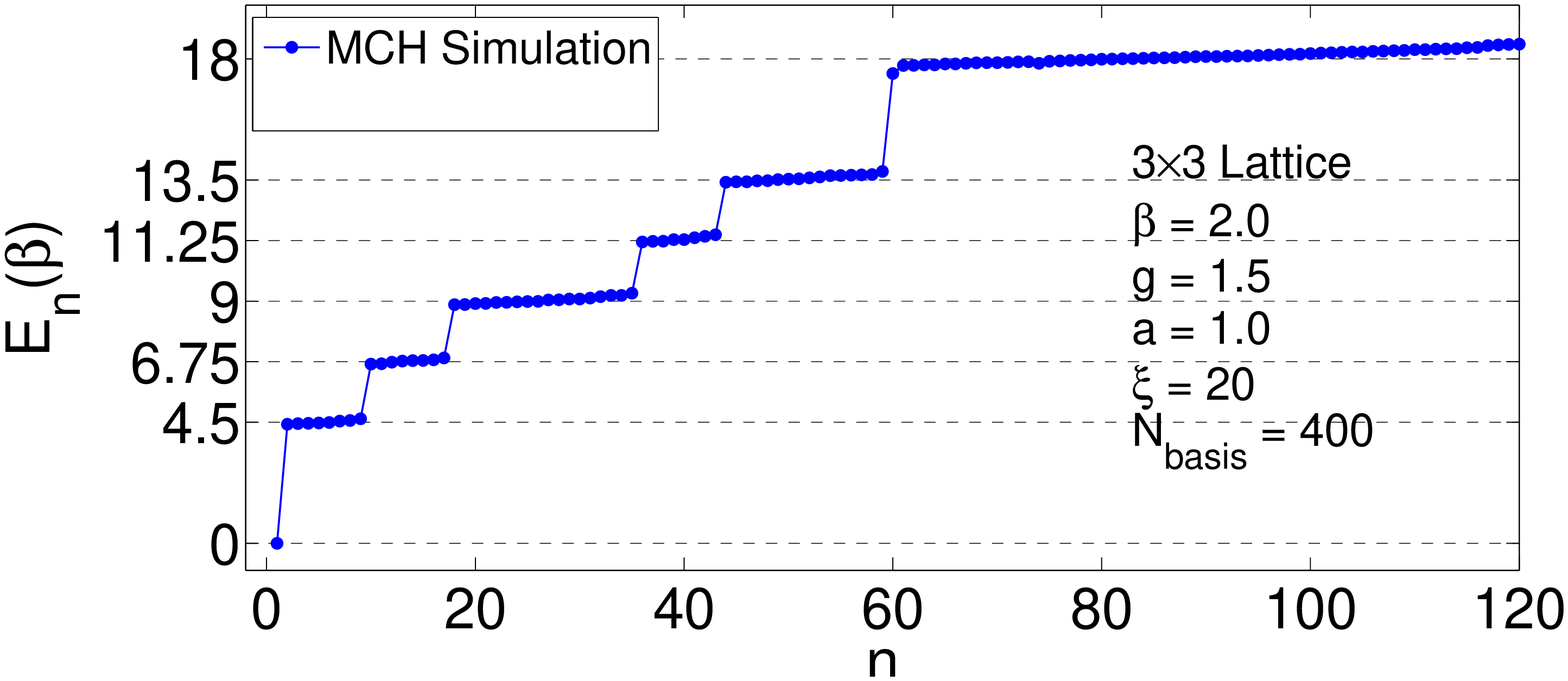}\\
 \includegraphics[width=85mm, height=40mm]{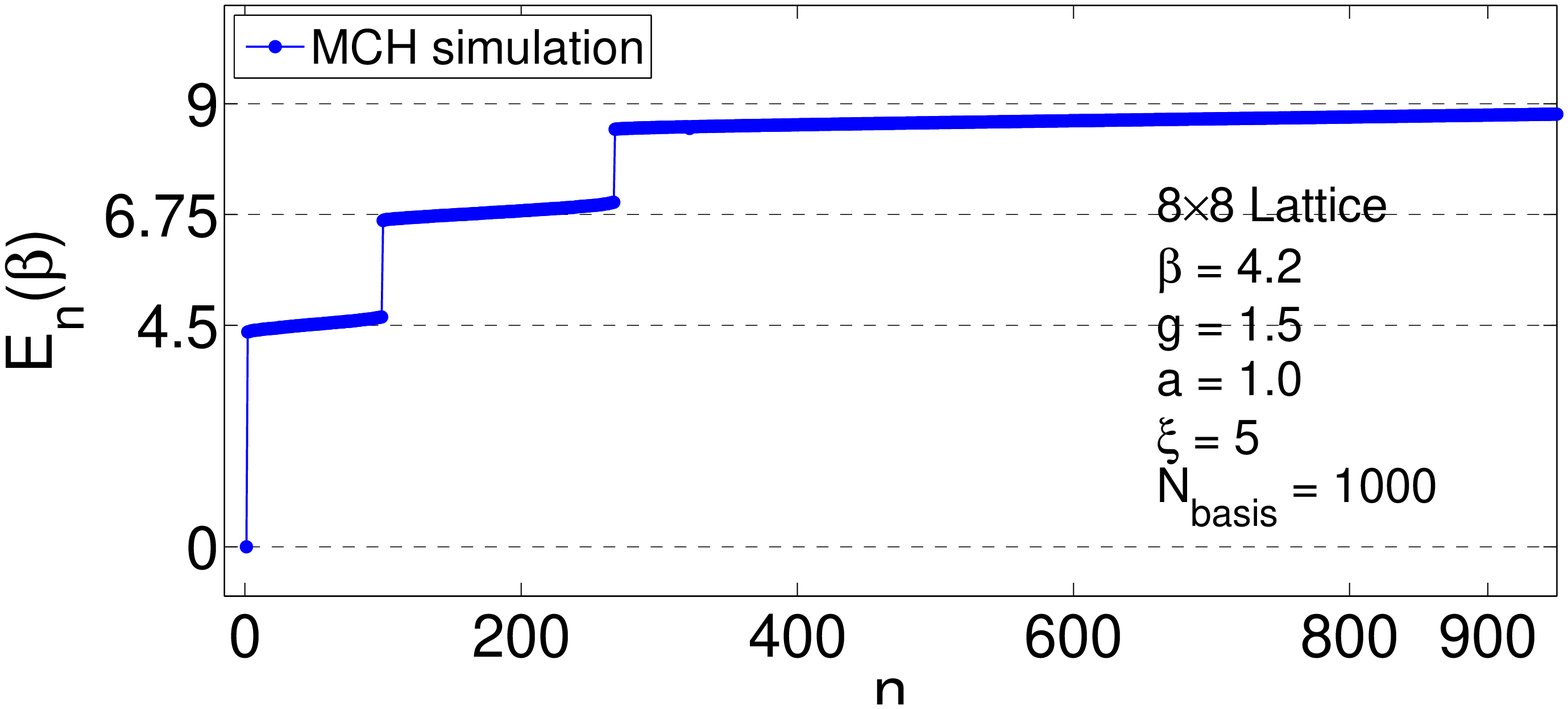}\\
 \includegraphics[width=85mm, height=40mm]{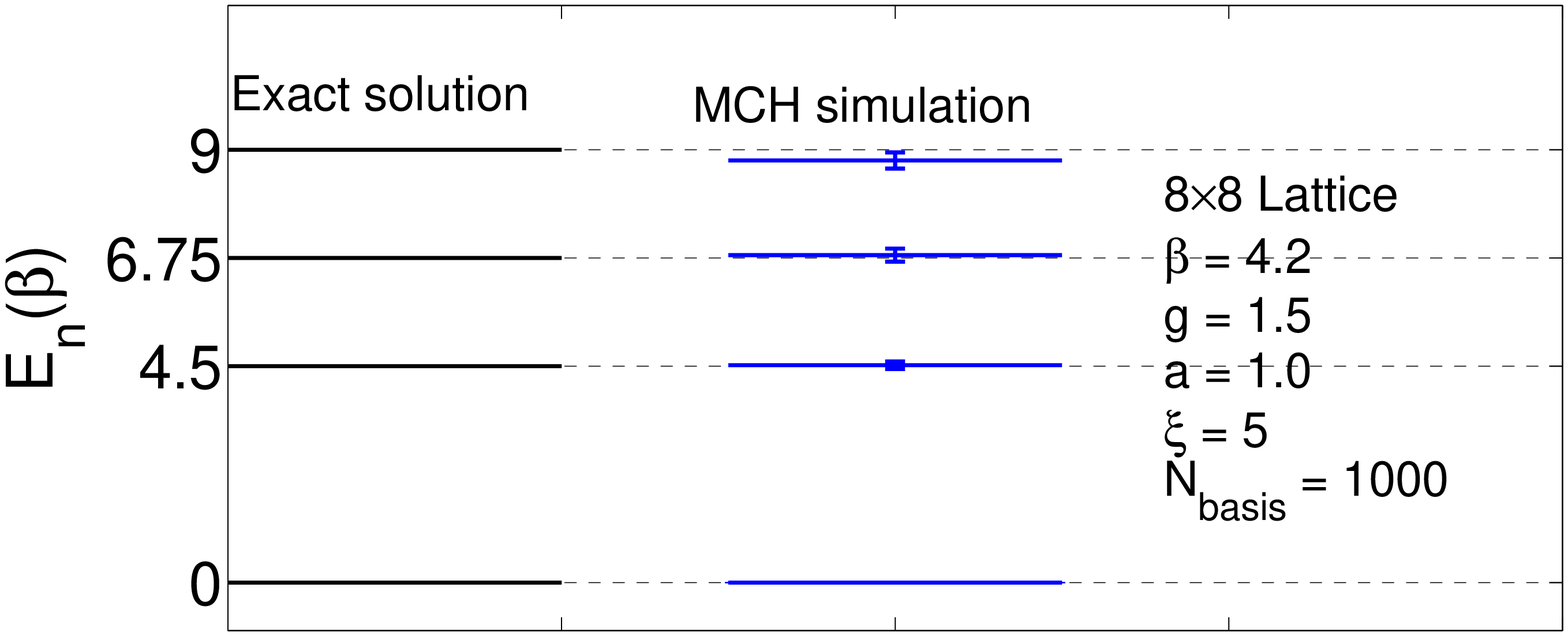}
 \caption{ Electric Hamiltonian. Eigenvalues of effective  Hamiltonian are compared to exact eigenvalues ($0$, $4.5$, $6.75$, $9.0$, $11.25$, $13.5$ and $18.0$). Top panel: spatial lattice $3^2$, imaginary time $\beta=2.0$, spacing $a=1$,  coupling constant $g=1.5$, asymmetric factor $\xi=20$, number of stochastic basis $N_\text{basis}=400$. Middle panel: lattice $8\times 8$, $\beta=4.2$, $a=1$, $g=1.5$, $\xi=5$, $N_\text{basis}=1000$. Bottom panel: The same results as the middle panel along with relative errors. Notice that the extension of horizontal parts in the top and middle panels indicates the eigenvalue degeneracy, which grows with the lattice size.}
 \label{fig:Spec1+2Plaq}
 \end{figure}
\begin{figure}[h]
\centering
 \includegraphics[width=90mm, height=45mm]{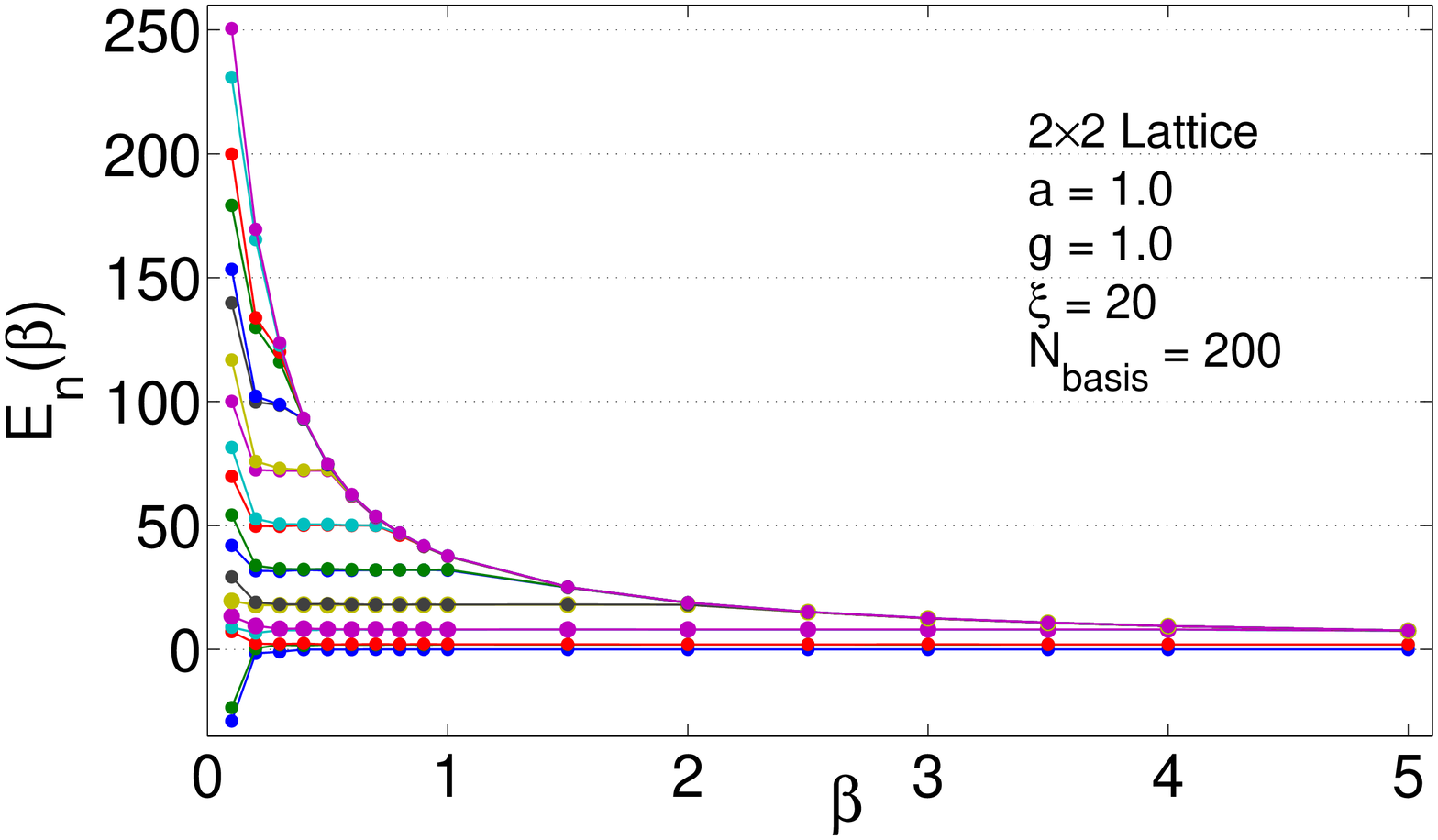}\\
 \includegraphics[width=90mm, height=45mm]{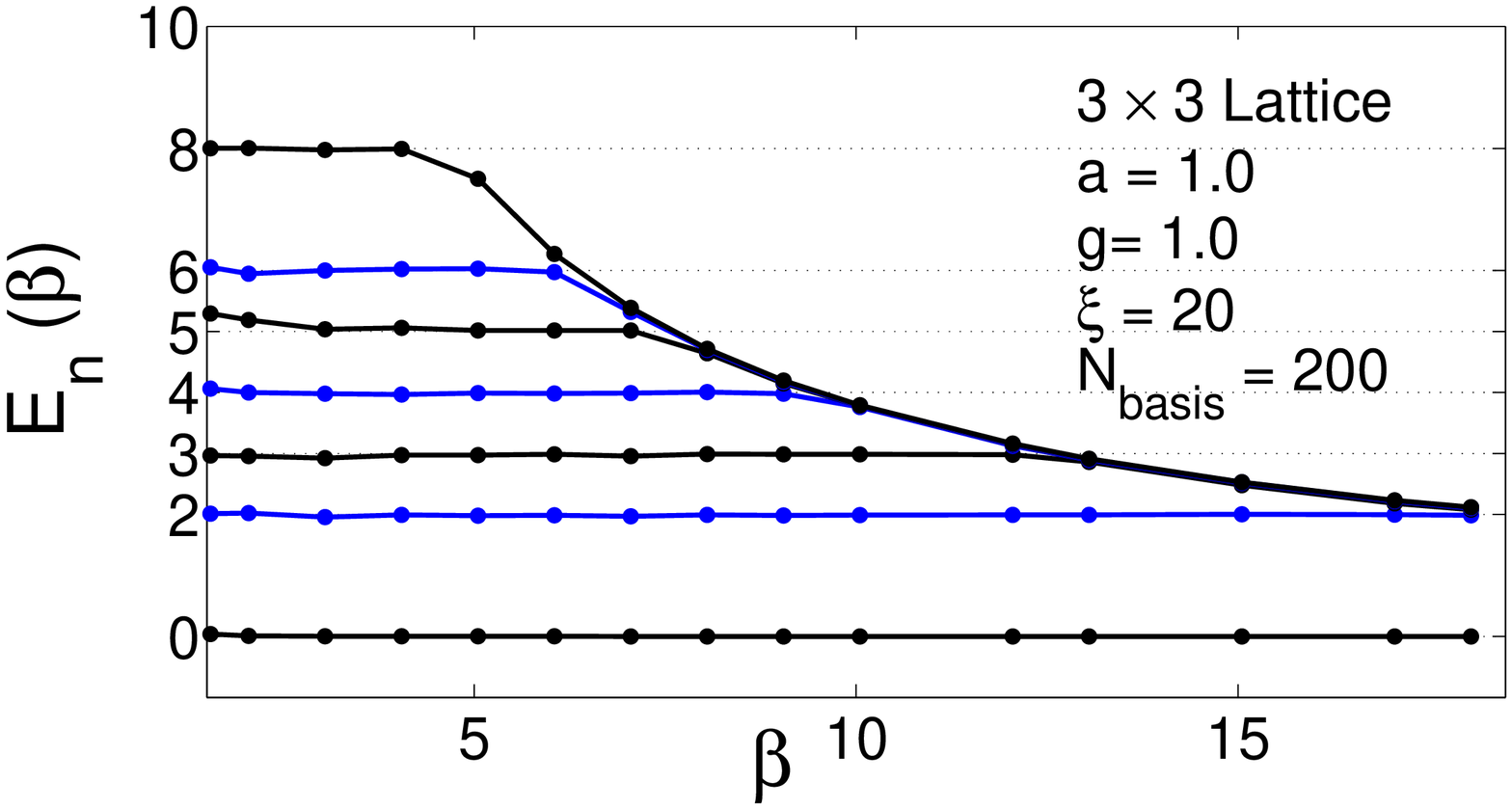}\\
 \includegraphics[width=90mm, height=45mm]{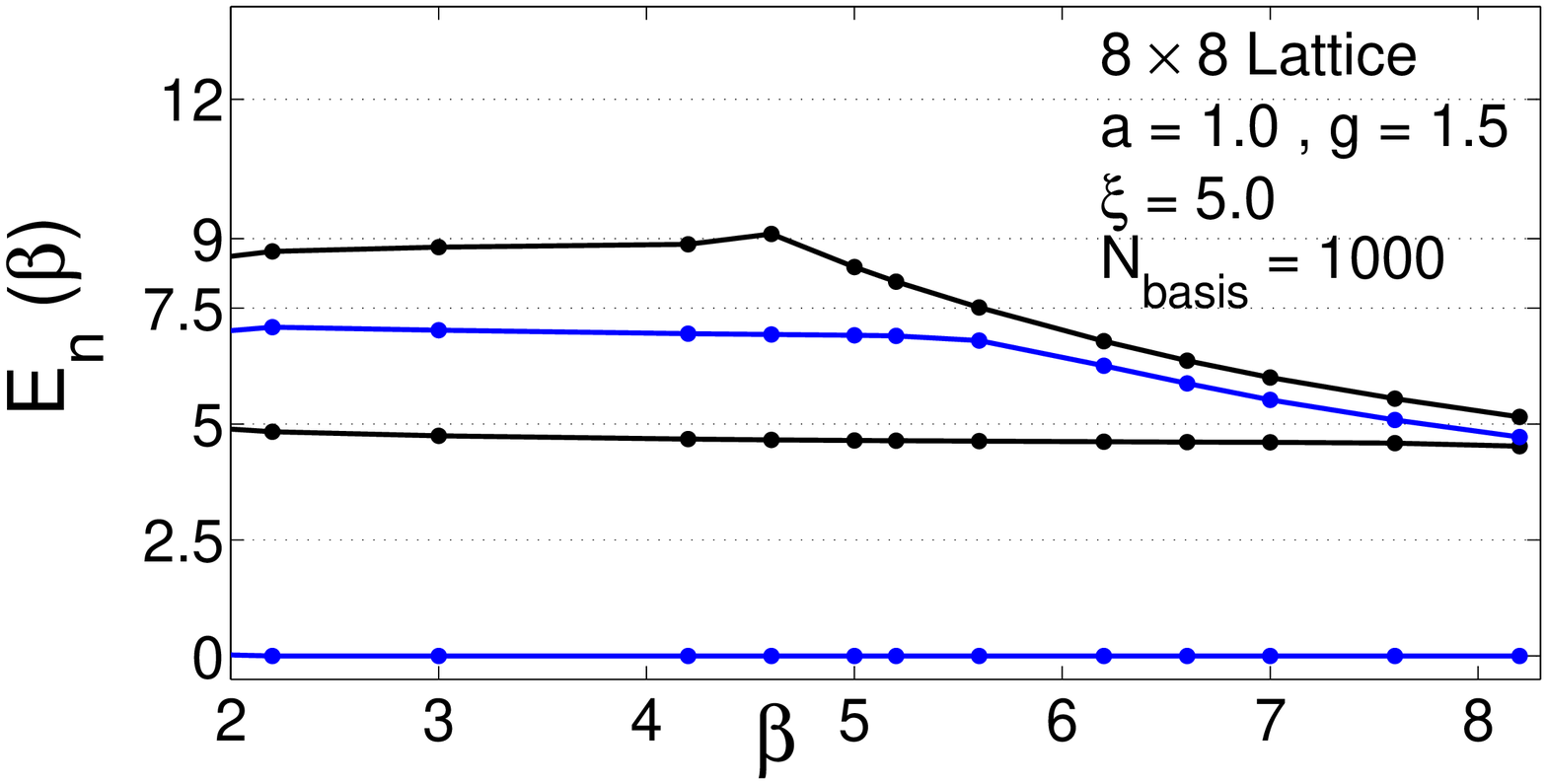}
\caption{Electric Hamiltonian. Scaling windows of energy eigenvalues of effective Hamiltonian are presented (the full line is a guide to the eye). The regime where the line is horizontal (energy scales in $\beta$ with exponent zero) represents a scaling window. Top panel: Spatial lattice $2^{2}$, $a=1$, $\xi=20$, $g=1.0$, $N_\text{basis}=200$. Middle panel: Spatial lattice $3^{2}$, $a=1.0$, $\xi=20$, $g=1.0$, $N_\text{basis}=200$. Bottom Panel: Spatial lattice $8^2$, $a=1$, $\xi=5$, $g=1.5$, $N_\text{basis}=1000$.}
\label{fig:FixedBasisScalingEnerg}
\end{figure}
The relative error stays in the order of $10^{-7}$ for the first 5 levels, and then grows exponentially with increasing level number $n$. The relative error reaches the order of 1 for $n=20$. Beyond that the spectrum of the effective Hamiltonian drowns in numerical noise. This means that there is an energy window, going from $E=0$ to $E=165$ where the effective Hamiltonian gives meaningful results. Such energy window is correlated with the behavior of diagonal elements $\mathcal D_{n}$ of the transition matrix $M$. The computations have been done with double precision, i.e., 15-16 digits. One observes that the upper bound of the energy window is reached, when the value of $\mathcal D_{20} \approx 0.81 \times 10^{-8}$ has 8 significant digits being exactly half of the 16 digits of internal arithmetic precision of the computer. The existence of such energy window has been observed in all cases. The size of the energy window depends on the following parameters:
(i) the internal numerical precision of the computer, and (ii) physical parameters, like coupling constant $g$, lattice size $a$, and transition time $T$. Notice that the transition time is equivalent to $\beta$, which is the inverse of temperature (in all figures we use $\beta$ instead of $T$). The internal numerical precision can be viewed as a scale of experimental observation, similar to the wave length of light in a microscope. The obtained results depend on the internal scale of the experimental apparatus, which in our case is the internal arithmetic precision of the computer. If we want to increase the size of the energy window we can (i) use higher numerical precision, which may be computationally costly, or (ii) choose physical parameters appropriately. 

One finds that larger lattices lead to an increased degree of degeneracy in the energy spectrum. Such degeneracy
comes from small closed loops, which located anywhere on the lattice, will give the same electric energy. Such high degeneracy will be lifted when taking the magnetic term into account. With increase of spatial lattice size an increase of the stochastic basis dimension is required. Also the tuning of time parameter $T$ in distribution $P(U)$, (Eq.~\ref{eq:DefProbDistr}) makes that $T$ changes with lattice size.
\begin{figure}[h]
\centering
 \includegraphics[width=90mm, height=45mm]{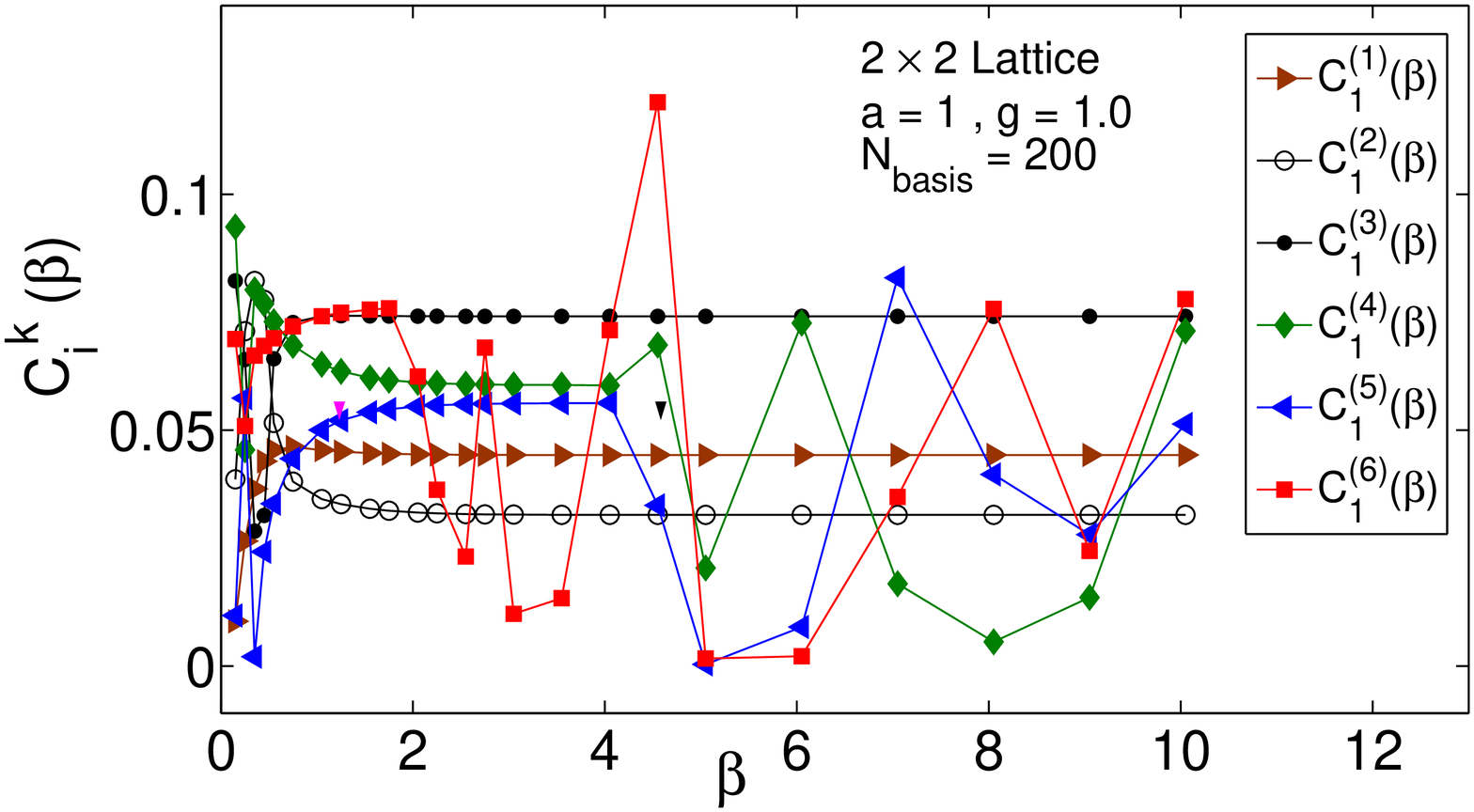}
\caption{Scaling of wave functions for the electric Hamiltonian. Expansion coefficients of wave functions corresponding to eigenvalues in Fig.~\ref{fig:FixedBasisScalingEnerg} (top panel) in terms of first stochastic basis. Spatial lattice $2 \times 2$, $a=1$, $\xi=20$, $g=1$, $N_\text{basis}=200$.}
\label{fig:FixedBasisScalingEnergWave}
\end{figure}
\begin{figure}[h]
\centering
 \includegraphics[width=90mm, height=40mm]{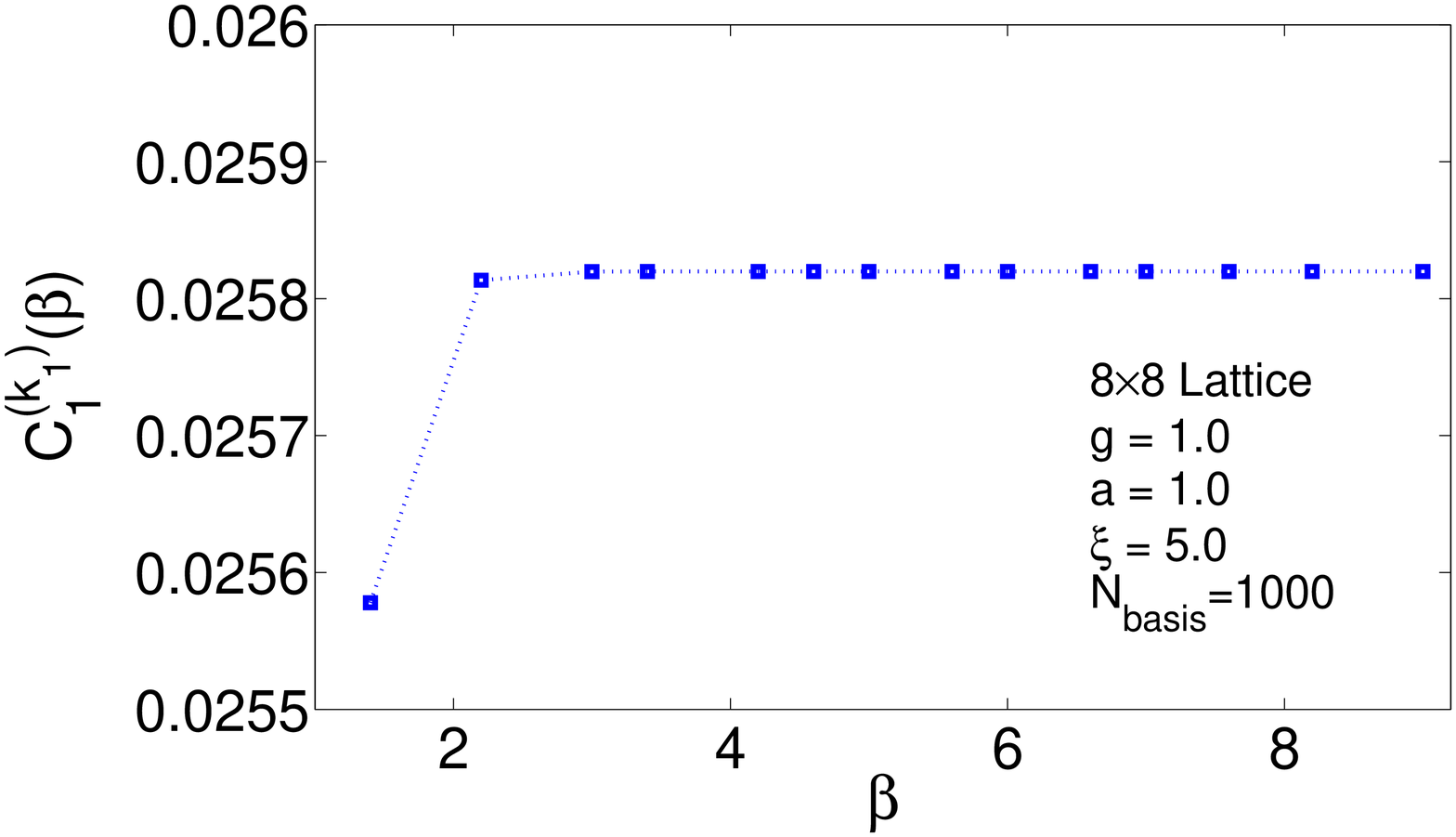}
 \includegraphics[width=90mm, height=40mm]{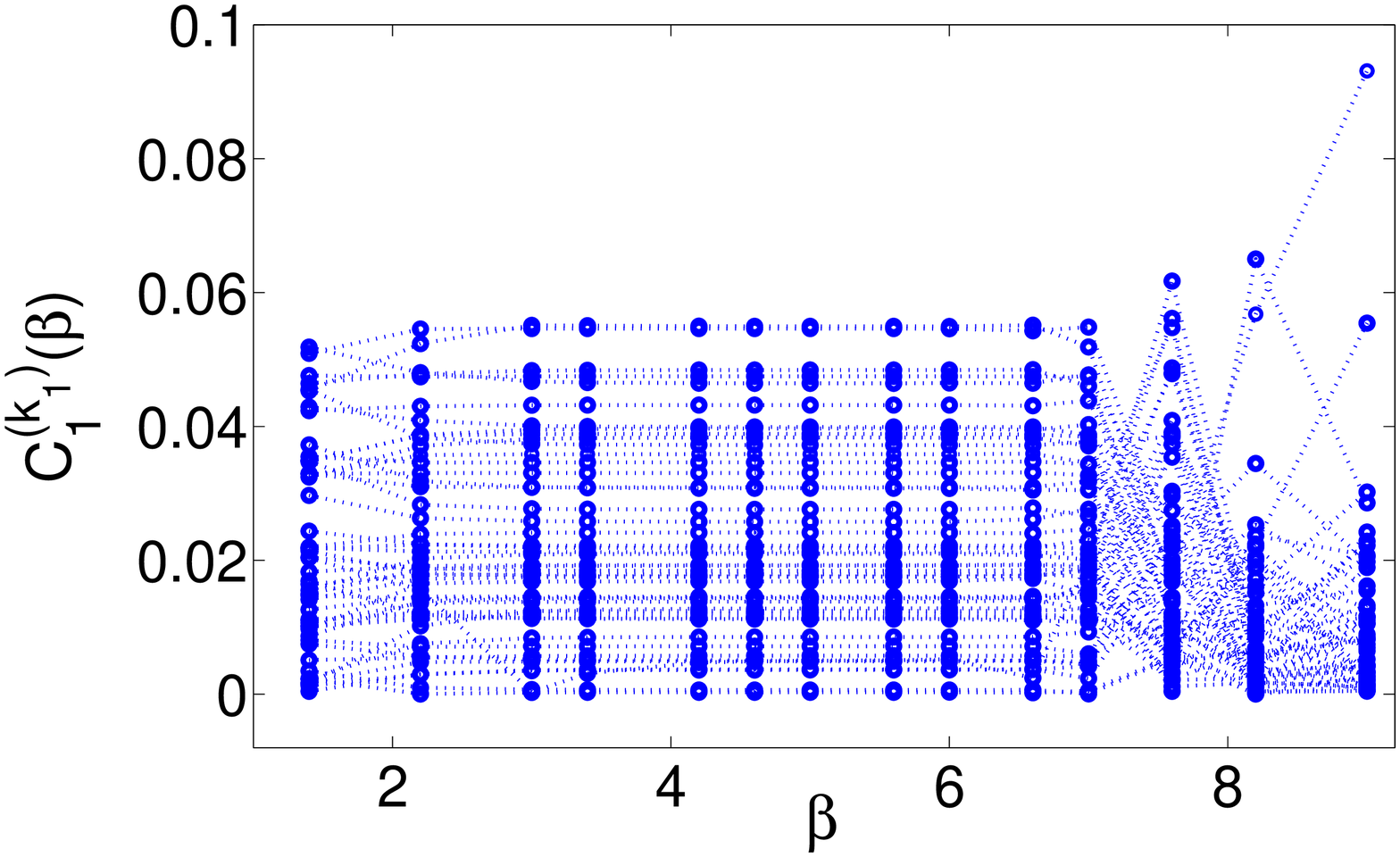}
 \includegraphics[width=90mm, height=40mm]{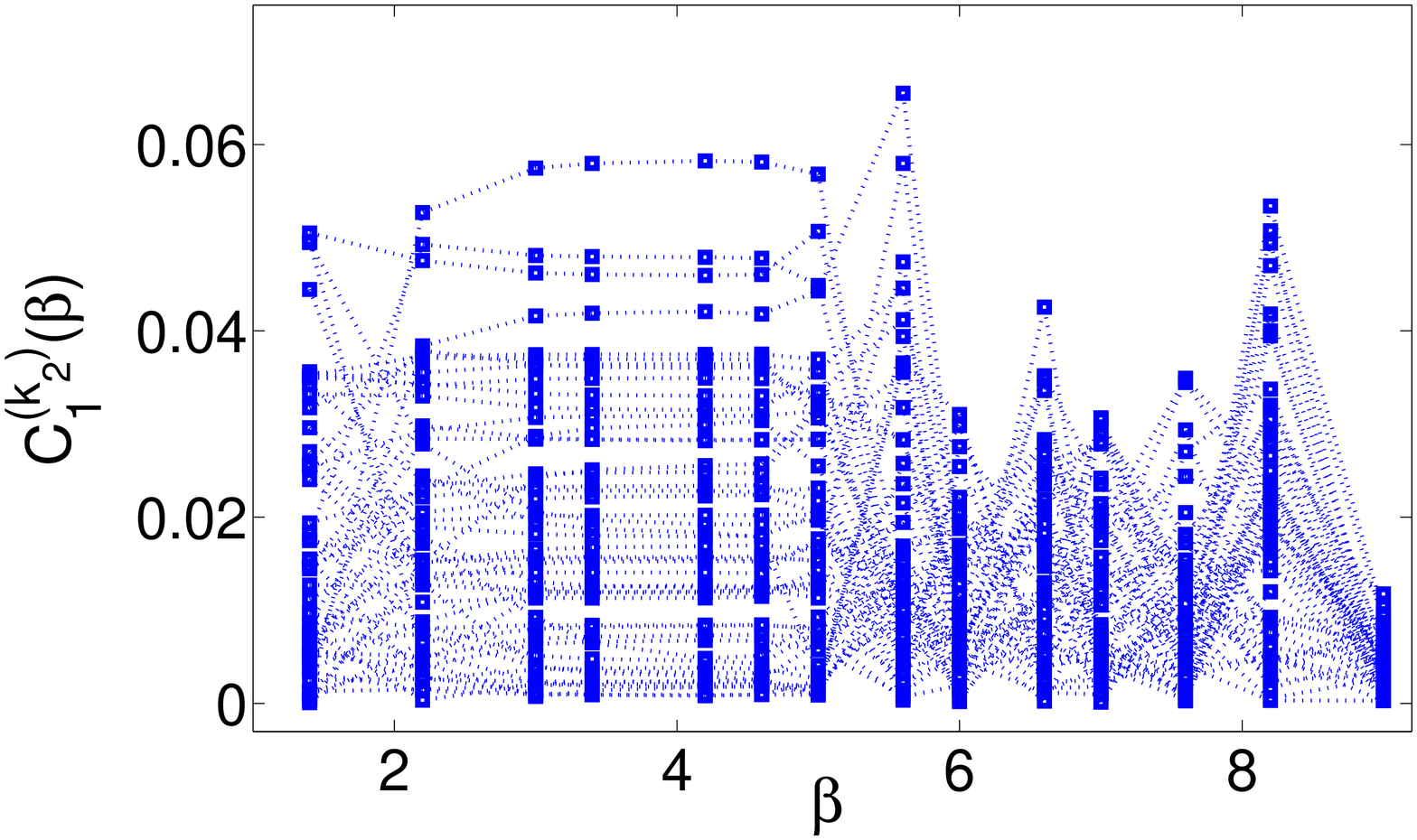}
 \includegraphics[width=90mm, height=40mm]{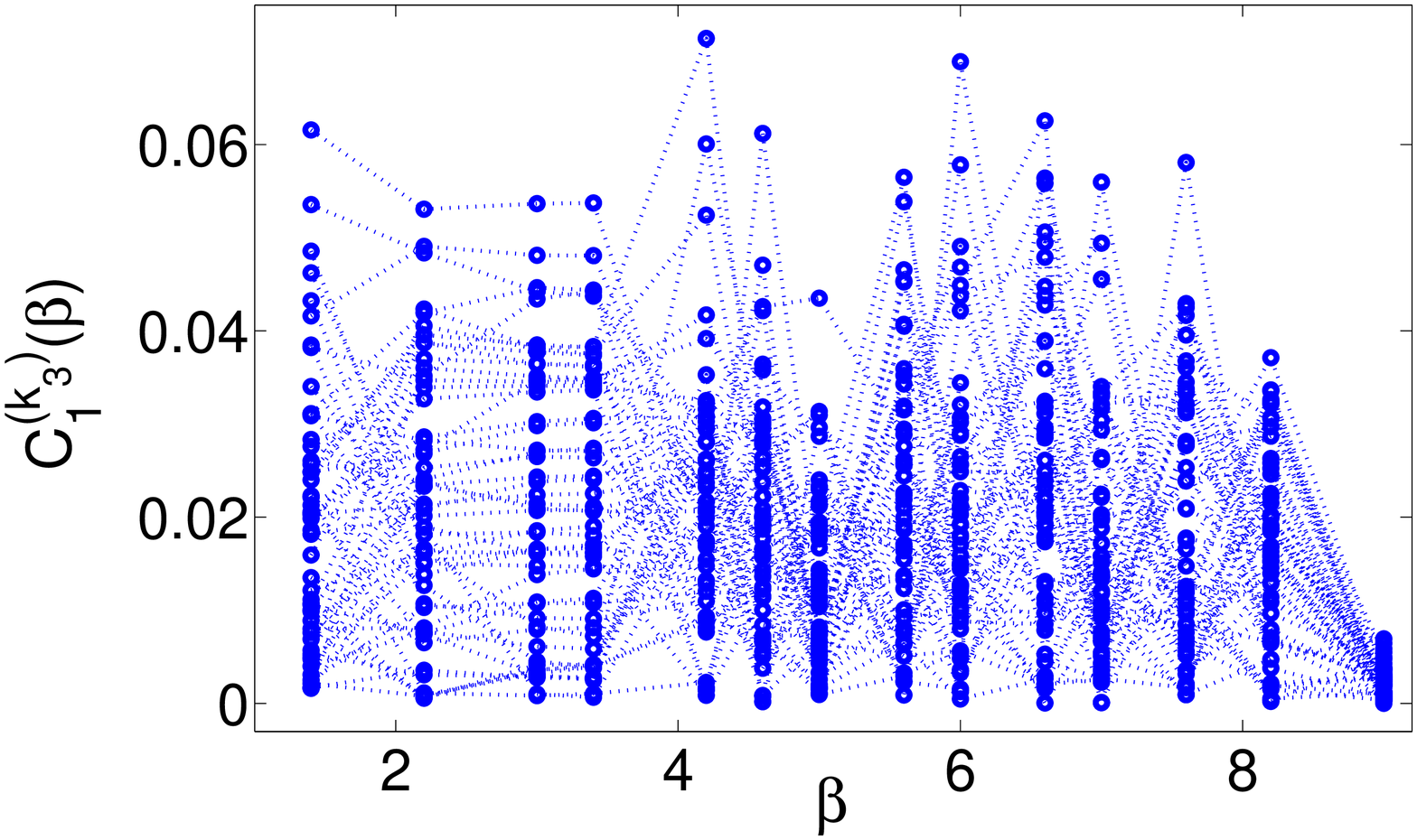}
\caption{Electric Hamiltonian. Scaling window of the expansion coefficients of wave functions, corresponding to eigenvalues in the bottom panel of Fig.~\ref{fig:FixedBasisScalingEnerg}. Spatial lattice $8 \times 8$, $a=1$, $\xi=5$, $g=1.5$, $N_\text{basis}=1000$. Notice that $k_1=1$ in the top panel relates to the vacuum state, and $k_2$, $k_3$, $k_4$ in the other panels are indices of the second, the third and the forth excited levels. Due to the high rate of degeneracy of these states, only some of them have been shown here.}
\label{fig:FixedBasisScalingEnergWave2}
\end{figure}

The above results all correspond to a particular value of transition time $T$ in the transition amplitude. However, the exact physical energy levels must be independent of the transition time $T$. The dependence of energy levels under variation of the parameter $\beta (\equiv T)$ has been investigated. Results are shown in Fig.~\ref{fig:FixedBasisScalingEnerg} for three  spatial lattices. The top panel (the $2^2$ lattice) shows that the ground state is a flat line as a function of $\beta$, in a wide range of $\beta$ (window from $\beta=0.2$ up to $\beta=5$), meaning that the ground state energy is independent of the transition time $\beta$. In other words, the ground state energy scales like $\beta^{\gamma}$ with exponent $\gamma=0$ in a finite time window. A similar scaling window is observed also for the higher excited states. However, with higher energy, the size of such scaling window becomes  smaller. For energy $E > 100$, the scaling window is no longer visible. We found that the size of the scaling window $S_{n}$ can be described approximately by an exponential law $S_{n} \propto \exp[-\sigma E_{n}]$. Such behavior of decreasing scaling windows can be understood from the property that $\exp[-\mathcal HT/\hbar]$ projects
onto the ground state for large $T$ (Feynman-Kac theorem). Higher levels become exponentially suppressed by the dominant ground state and can survive only for short times $T$.  The middle and the bottom panels of Fig.~\ref{fig:FixedBasisScalingEnerg} also display the scalling window of $3^{2}$ and $8^2$ spatial lattices. One observes that scaling windows are observed in a spectrum with higher degeneracy.
For the $8^2$ lattice the distribution $P(U)$ is expressed via the path integral, while for the $2^2$ and $3^2$ lattices it is given by the analytic expression (see Eq.~\ref{eq:AnalyticFormulae}).

The scaling behavior observed in the energy eigenvalues should be manifest also the corresponding wave functions. In particular, we have studied $\langle e_{\mu} | \Phi_{n} \rangle$, i.e. the expansion coefficient of wave function $\Phi_{n}$ in terms of the stochastic basis function $e_{\mu}$. 
As two examples, such expansion coefficients, expanded in terms of the first basis function, are shown for the lattice sizes $2^2$ and $8^2$ in Fig.~\ref{fig:FixedBasisScalingEnergWave} and Fig.~\ref{fig:FixedBasisScalingEnergWave2}, respectively. Indeed, one finds scaling behavior also in such expansion coefficients. Like the size of the energy scaling window decreases with increasing energy $E_{n}$, also the wave function has a scaling window, and its size also decreases with the level index $n$ of energy. Moreover, the eigen function degeneracy is observed for the $8^2$ lattice, and it is increased with the lattice size.
\begin{figure}[h]
\centering
 \includegraphics[width=90mm, height=40mm]{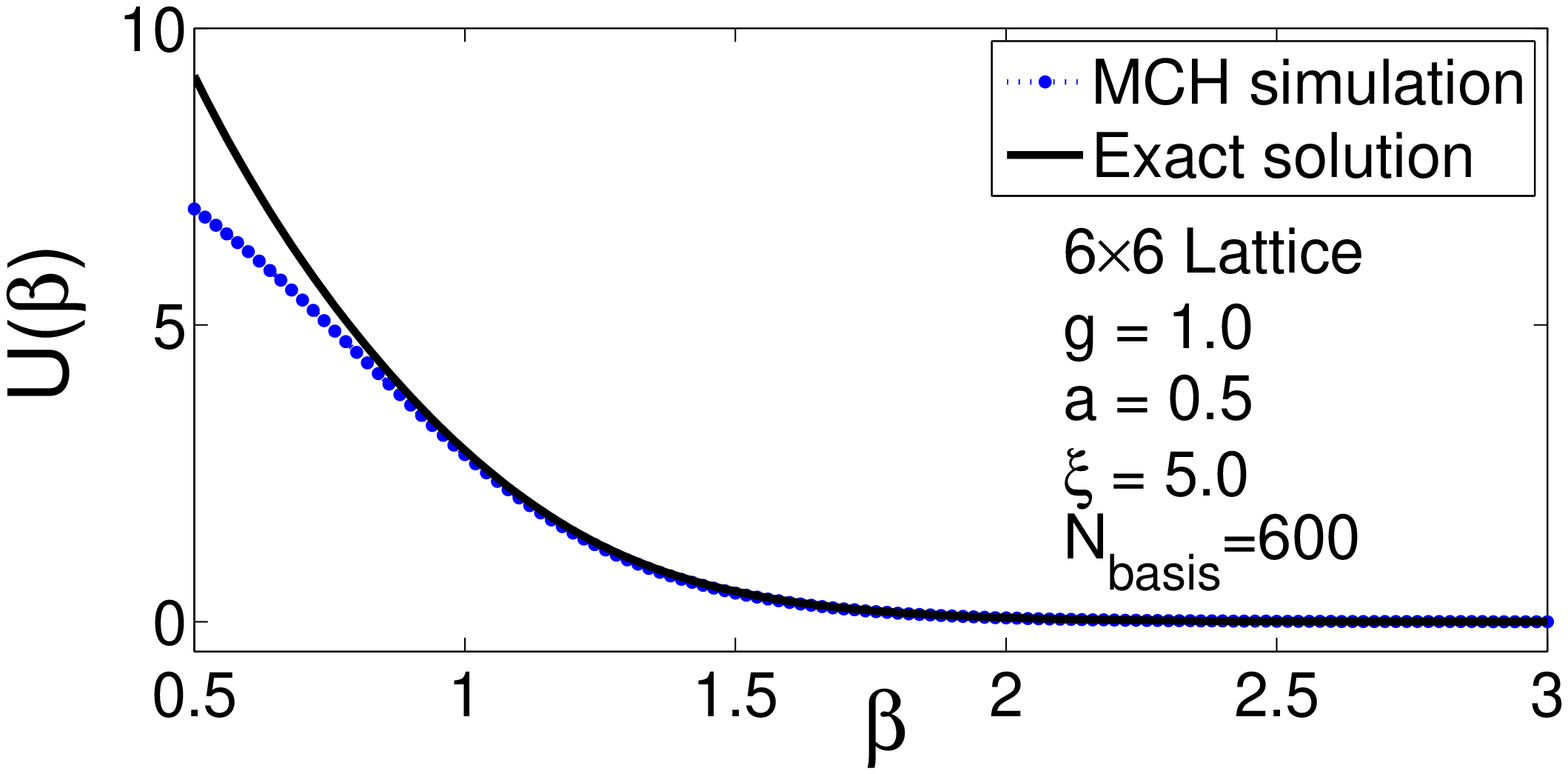}\\
 \includegraphics[width=90mm, height=40mm]{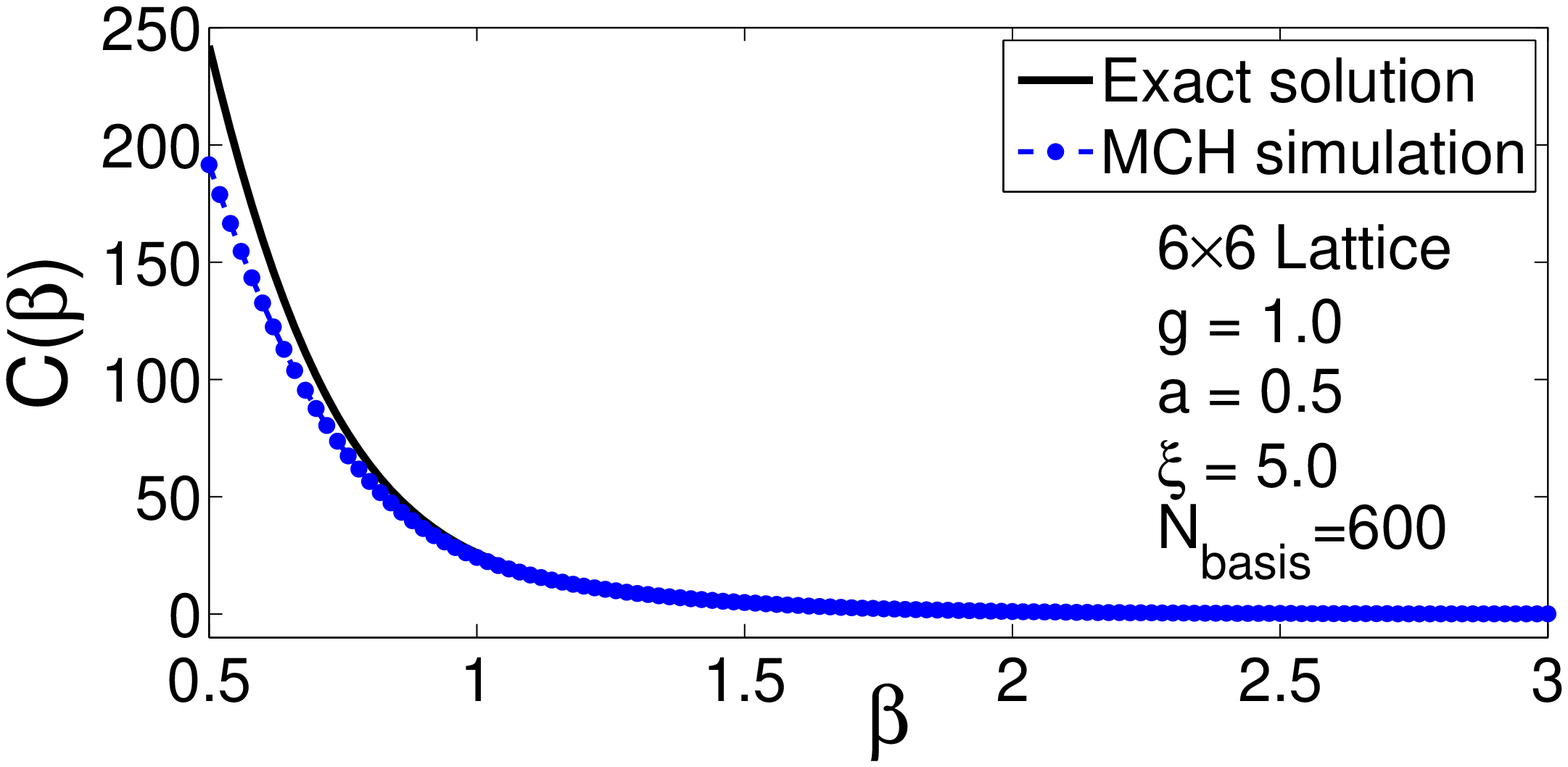}\\
\caption{
Thermodynamical functions computed from exact eigenvalues (full line) versus those from eigenvalues of effective electric Hamiltonian for a $6\times6$ lattice with $a=1$, $\xi=5~(a_0=0.2)$, $g=1.0$ and $N_\text{basis}=600$. Top panel: Average energy $U(\beta)$. Bottom panel: Specific heat $C(\beta)$. In lower temperatures (higher values of $\beta$) there is a reasonable agreement.
}
\label{fig:ThermodynFixedBasis}
\end{figure}

{\it Thermodynamics.}
The quality of the energy spectrum of the effective Hamiltonian can also be seen from the thermodynamical functions. In the case of the electric Hamiltonian, the energy spectrum $E_{0}, E_{1},\dots$ can be computed analytically. Thermodynamical functions can be expressed via those energies. The partition function $Z$ is denoted by
\be
Z(\beta) = Tr\left[\exp(-\beta\mathcal H)\right]= \sum_{n} \exp\left[- \beta E_{n}\right ] ~ ,
\ee
and the average energy $U(\beta)$, e.g., is given by
\be
U(\beta) = - \frac{\partial \log Z(\beta)}{\partial \beta}=\frac{1}{Z(\beta)}\sum_{n} E_n \exp\left[-\beta E_n\right]~ .
\ee
Likewise, one can compute free energy, entropy and specific heat. 
For example, Fig.~\ref{fig:ThermodynFixedBasis} shows  the results of the average energy $U(\beta)$ and the specific heat $C(\beta)$ for the electric Hamiltonian of a $6^2$ lattice. In this figure we have compared the results from the effective Hamiltonian with the exact Hamiltonian. In general one observes that agreement is good in the regime of large $\beta$, i.e., the low temperature regime. For small values of $\beta$ some disagreement becomes visible, reflecting the fact that the precision of higher energy levels of the effective Hamiltonian is limited (their scaling windows go to zero).

%
\section{Random perturbation of electric Hamiltonian}
\label{sec:RandPert}
\begin{figure}[h,t]
\centering
 \includegraphics[width=90mm, height=40mm]{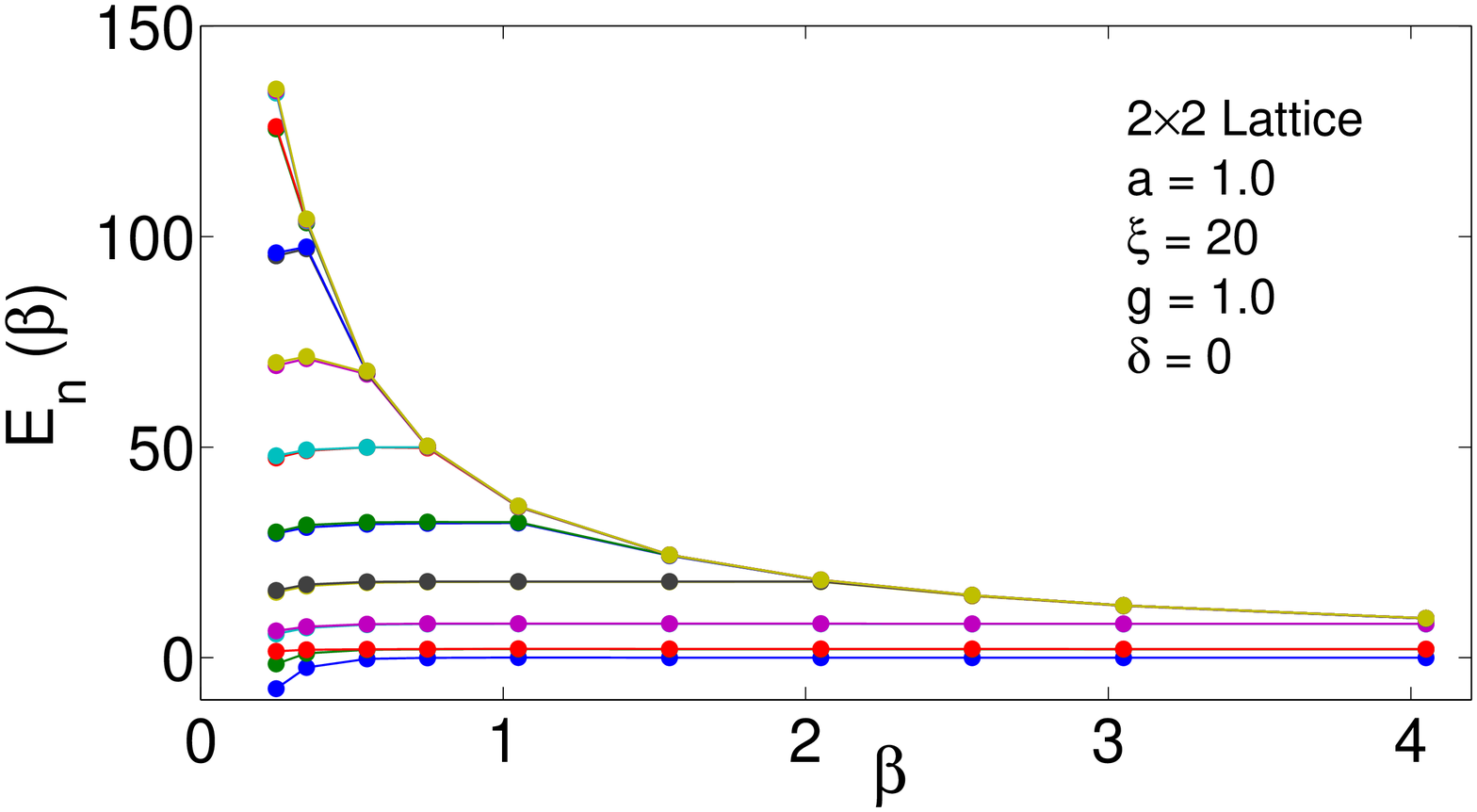} \\
 \includegraphics[width=90mm, height=40mm]{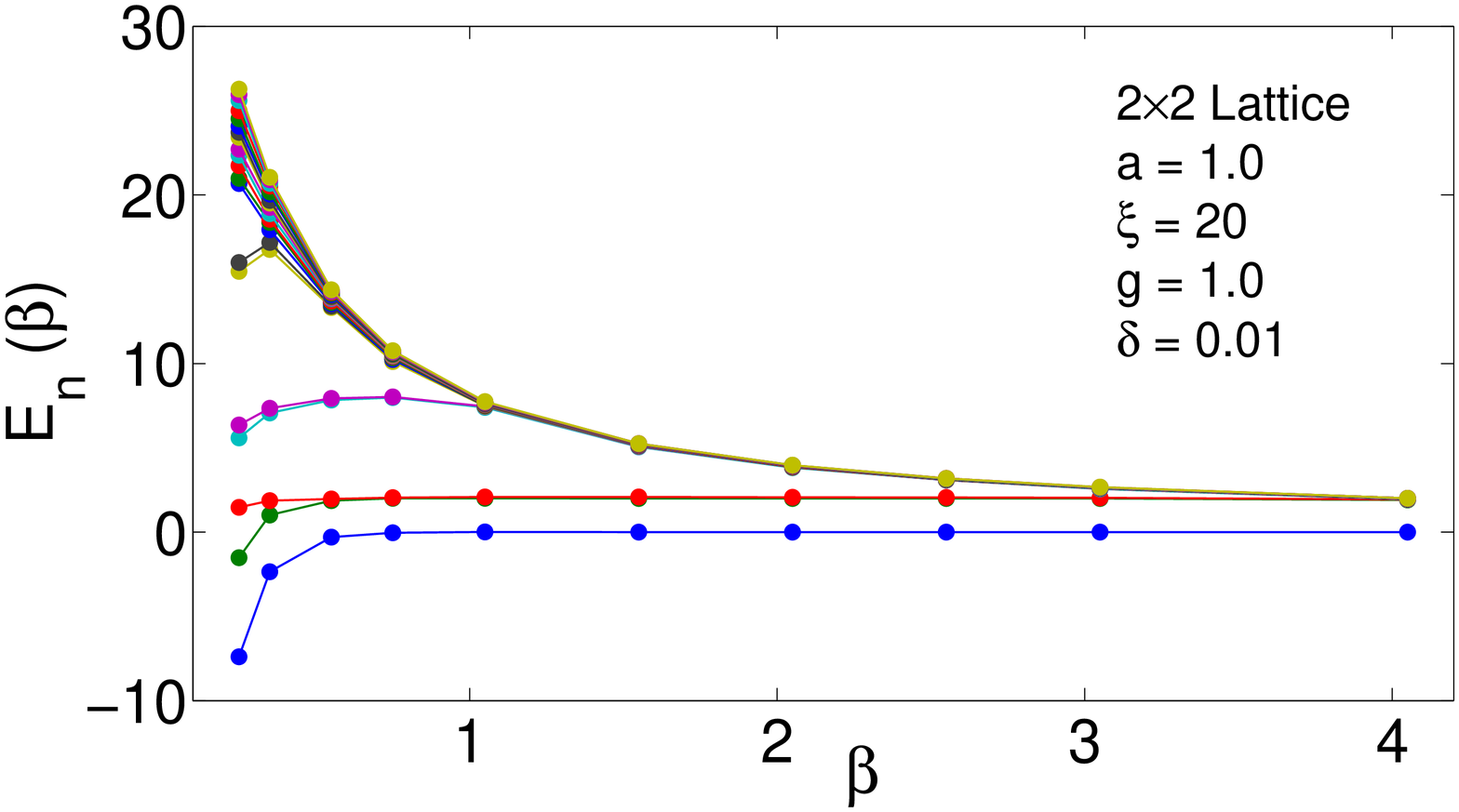} \\
 \caption{Electric Hamiltonian. Influence of artificial random relative error $\delta$ in transition matrix on
 scaling behavior of energies. A $2^{2}$ spatial lattice. Top panel: $a=1$, $\xi=20$ (lattice asymmetry parameter), $g=1$, $N_\text{basis}=200$, $\delta=0.01$ (no error).
 Bottom panel: the same graph with error ($\delta=0.01$).} 
\label{fig:RandErrInflScaling}
\end{figure}

%
Below we will present results for the effective Hamiltonian of the full Hamiltonian. In order to better understand the propagation of numerical errors, we have considered the electric Hamiltonian plus a small random perturbation, which shall simulate numerical errors. We have then computed the spectrum of the corresponding effective Hamilton, and compared the cases with and without the random perturbation. The results are shown in Fig.~\ref{fig:RandErrInflScaling}. One clearly observes that the random perturbation reduces the number of states which show scaling and also reduces the size of the scaling windows.
%
%

%
\section{Transition amplitudes under full Hamiltonian}
\label{sec:TransAmpl}
%
Let us consider the gauge invariant transition amplitude under the full Hamiltonian (in analogy to quantum mechanics (see Eq.(\ref{eq:TransAmplQM})). Although this can be expressed in terms of a path integral with the lattice action (Wilson action), this is numerically not suitable, because Monte Carlo with importance sampling only allows to compute ratios of transition amplitudes. Hence, like in quantum mechanics (Eq.(\ref{eq:TransAmplPathInt})) we factorize the above amplitude into two terms, one being analytically computable and the other one being given by the ratio of transition amplitudes computable via the method of Monte Carlo.
Therefore, the transition amplitude under evolution of
the full Hamiltonian, between gauge invariant projected states is written as
\bea
\label{eq:MatrixElemGauge}
&& M_{\mu,\nu}(T)
 = \langle U_{\mu} |\hat{\Pi} ~ \exp[-\mathcal H_\text{full} T/\hbar ] | U_{\nu} \rangle
\nonumber \\
&& 
\hspace{1.2cm} = \langle U_{\mu} |\hat{\Pi} ~ \exp[-\mathcal H_\text{elec} T/\hbar ] | U_{\nu} \rangle
\nonumber \\
&& \times
\frac{
\left. \int [dU] ~ \exp[ - S[U]/\hbar ] \right|^{U_{\mu},T}_{U_{\nu},0}
}
{
\left. \int [dU] ~ \exp[ +S_\text{mag}[U]/\hbar ] \exp[ -S[U]/\hbar ] \right|^{U_{\mu},T}_{U_{\nu},0}
} ~ .
\nonumber \\
\eea
%
%
Here $U_{\mu}$ denotes the stochastic basis of Bargmann states on the lattice (in analogy to the quantum mechanical stochastic position states $x_{\mu}$). We have drawn these states from the distribution corresponding to the electric Hamiltonian, given by Eq.~(\ref{eq:DefProbDistr}).
We have preferred this choice because this function is analytically computable, which means a numerical effort being substantially smaller than that for the distribution involving the full Hamiltonian, Eq.~(\ref{eq:TrialProbDistr}). We construct a matrix of transition elements between normalized stochastic basis states. We proceed in analogy to quantum mechanics (Eq.~\ref{eq:ApproxTransAmplBoxStates}). Finally, we diagonalize such matrix and extract eigenvalues and wave functions (Eqs.~\ref{eq:DiagonalMatrix}-\ref{eq:EnergEigValue}).
%
%
\begin{figure}[h,t]
\centering
 \includegraphics[width=90mm, height=40mm]{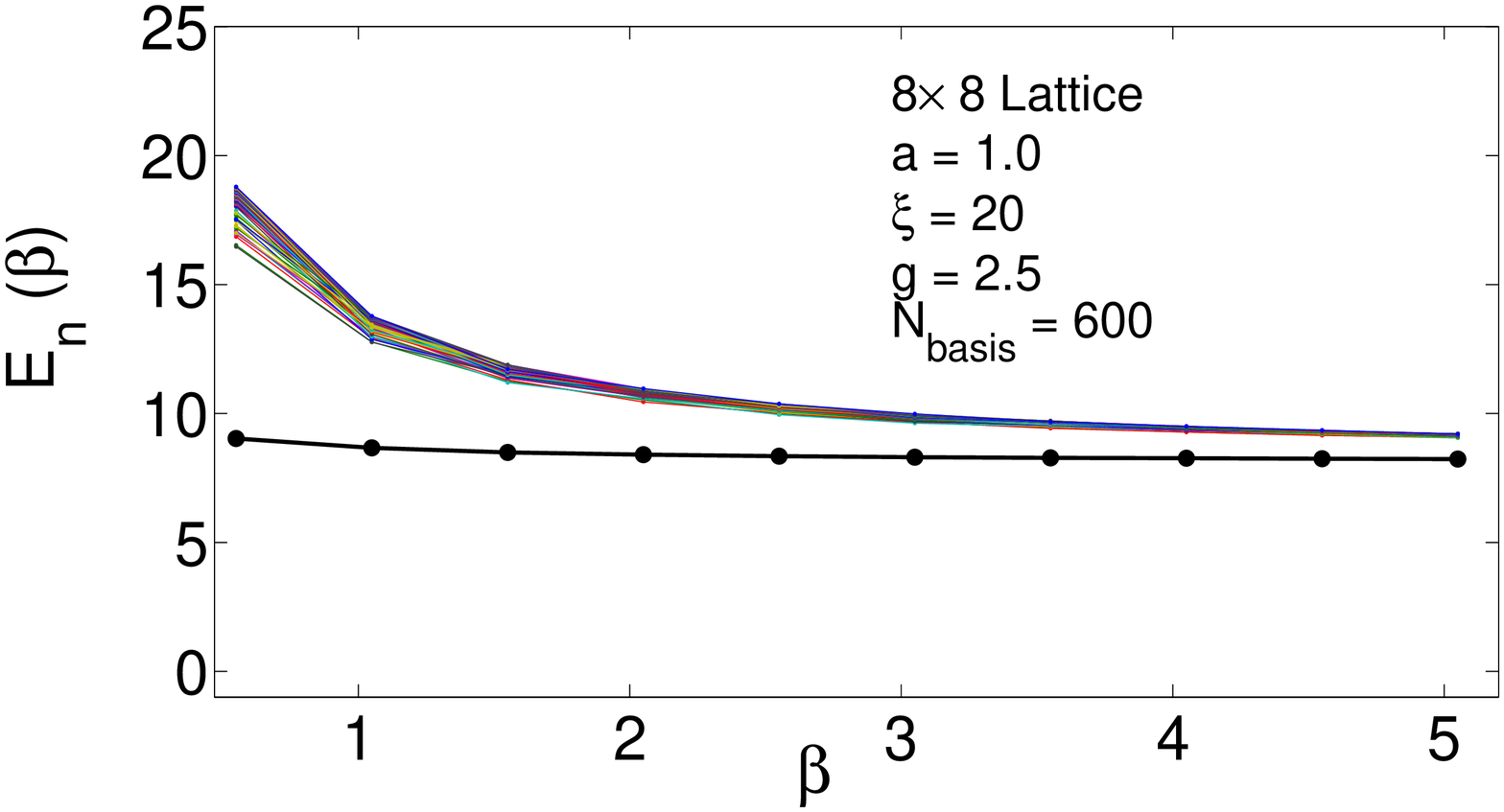} \\
 \includegraphics[width=90mm, height=40mm]{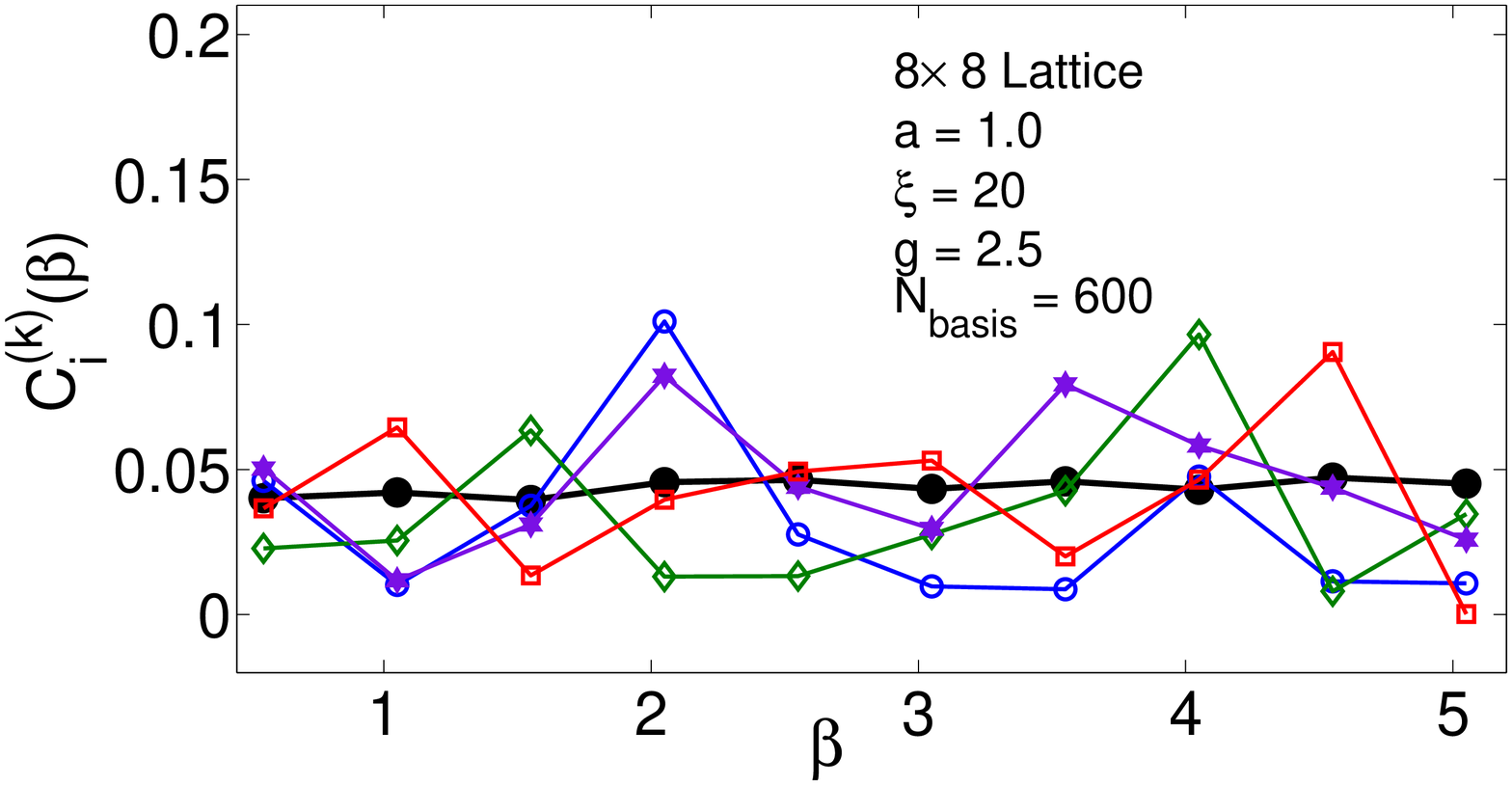}
\caption{Full Hamiltonian: Scaling window of the spatial $8 \times 8$ lattice with $a=1$, $g=2.5$, $\xi=20$, $N_\text{basis}=600$. To panel: ground state energy. Bottom panel:  expansion coefficient of the ground state wave function.}
\label{fig:ScalingCompareElec+FullHam}
\end{figure}

{\it Full Hamiltonian: influence of the magnetic term}. Taking into account the magnetic term generates the effective full Hamiltonian.
First results on scaling of energy eigenvalues and wave functions of low-lying states are shown in Fig.~\ref{fig:ScalingCompareElec+FullHam}. These results correspond to a $8^2$ lattice. The results show scaling windows for the ground state energy (top) and the expansion coefficient of the ground state wave function (bottom).
Comparing with the scaling observed in the electric Hamiltonian, only one level shows scaling,
and the scaling window is smaller. This can be understood from the fact that the ratio of matrix elements in Eq.~(\ref{eq:MatrixElemGauge}) has been determined via Monte Carlo from path integrals, which carries statistical errors in the order of a few percent. We have seen that errors of such order of magnitude de reduce the scaling behavior (see Fig.~\ref{fig:RandErrInflScaling}). From this observation we conclude that the numerical resolution of
energy levels and scaling windows in the full Hamiltonian is essentially determined by the statistical error occurring in the numerical calculation of the ratio of matrix elements. Results with better statistics and a larger stochastic basis are being called for.

%
\section{Discussion}
\label{sec:Discuss}
%
The construction of a Hamiltonian in lattice gauge theory faces the following problems:

(a) Although there exists the lattice Hamiltonian, this alone has not proven to be useful in computing viable results for physical observables.
We suggest here that the best one can do is to construct an effective Hamiltonian. This is meant to be a Hamiltonian which describes physics in a finite window (e.g. a window of low energy). This has some analogy to the idea of of the Wilson-Kadanoff renormalisation group, where a renormalized Hamiltonian is constructed, which is valid at some critical point, but distant from the critical point has no physical meaning.

(b) Conventional Hamiltonian methods used in atomic physics, nuclear physics, condensed matter and particle physics consider the given Hamiltonian $\mathcal H$ and compute matrix elements $\langle \phi_{i} | \mathcal H | \phi_{j} \rangle$ from such Hamiltonian. In contrast, here we consider a function of the Hamiltonian $\exp[ - \mathcal H T/\hbar]$. This has the following advantages: First, $\mathcal H$ is mathematically a more singular object than $\exp[-\mathcal H]$. This can be seen at hand of the simple quantum mechanical example of the kinetic Hamiltonian $\mathcal H_\text{kin}$. The matrix element $\langle y |\mathcal H_\text{kin} | x \rangle$ is a derivative of a $\delta$-function, while the matrix element $\langle y |\exp[-\mathcal H_\text{kin}] | x \rangle$ is a smooth, differentiable and rapidly falling-off function. This means, the non-linear exponential function smoothes out singularities of an operator. Second, matrix elements of the exponential function $\exp[ - \mathcal H T/\hbar]$ can be
evaluated by using the path integral. Third, and mostly important, in contrast to $\mathcal H$, the operator function $\exp[ - \mathcal H T/\hbar]$ contains a parameter $T$, which is redundant for the physical spectrum, i.e. using any value of $T$ one should obtain the same spectrum of $\mathcal H$. In numerical simulations such redundancy gives additional information about errors in the following way: The results do depend on the value of $T$. They depend also on physical parameters, such as coupling, lattice size and lattice spacing etc. They further depend on approximation parameters, like the number of equilibrium configurations used in the path integral, the size of the stochastic basis etc. Last but not least they depend on the internal precision used in the computer. Here we can turn the dependence on $T$ into an advantage: (i) The $T$-dependence of the energy spectrum (or better of a number of low lying energy eigenvalues) can serve as a measure of error of the calculation. In the best case the energy eigenvalues become $T$ independent. This happens in the so-called scaling windows. In the worst case they are strongly $T$-dependent, meaning that these results are unphysical. (ii) We can use the scaling window to tune the time parameter occurring in the distribution $P(U)$, which generates the stochastic basis.

(c) Stochastic basis. In our opinion, the construction of suitable basis is the most important step in order to compute physics from a Hamiltonian.
Such basis is built on two principles: first a random pick and second, a physical principle to guide the search. The so-called stochastic basis is
built in close analogy to the equilibrium path configurations computed via Monte Carlo importance sampling to solve Lagrangian lattice gauge theory.

(d) In contrast to Lagrangian lattice gauge theory, where gauge symmetry is manifestly conserved  in the path integral via the group measure and
the (Wilson) action, in the Hamilton formulation gauge invariance of states and amplitudes has to be imposed (via Gauss' law). Technically, much work
is required to construct such gauge invariant states. One expands the link Bargmann states into irreducible representations using the Peter-Weyl theorem. Gauge invariant states are then constructed by doing the group integral of local gauge transformations (at each vertex). As a result one enforces Gauss' law at each vertex. Here, we have shown how this can be done in the case of U(1) gauge theory. This can be generalized to non-Abelian gauge symmetry. For example, Burgio et al.~\cite{Burgio00} have shown how to construct a gauge invariant Hilbert space for the gauge group SU(2).

\vspace*{6pt}

\noindent {\bf Acknowledgement.} H. Kr\"oger and M. McBreen have been supported by NSERC Canada. This paper is dedicated to the memory of Prof. X.Q. Luo.

\vspace*{6pt}
%

%
\section{Appendix A: Peter-Weyl theorem for SU(N)}
\label{Appendix}
%
We consider the Lie group SU(N).
Irreducible representations play a crucial role in the Peter-Weyl theorem in the case of U(1) and also for SU(2). This is also true in general for SU(N). Let us consider irreducible representations in the notation of Young diagrams characterized by a partition and denoted by 
\begin{equation}
\label{eq:IndexIrredRep}
\{\nu\} \equiv \{\nu_{1};\dots;\nu_{N-1}\} ~ , \mbox{where} ~  
\nu_{1} \ge \dots \ge \nu_{N-1} ~ . 
\end{equation}
It is convenient to define the number 
\begin{equation}
\bar{\nu}=\sum_{i=1}^{N-1} \nu_{i} ~ .
\end{equation}
Elements of the group manifold can be conveniently parametrized by
\begin{equation}
\label{eq:GroupElem}
\mathscr G = \exp\left[ i \sum_{j=1}^{N^{2}-1} \hat{\tau}_{j} \phi_{j}/2 \right] ~ ,
\end{equation}
where $\hat{\tau}_{1}, \dots, \hat{\tau}_{N^{2}-1}$ denote the group generators in the fundamental representation. A group element $\mathscr G$ in the irreducible representation characterized by index (set) $\{\nu\}$ is given by a matrix
\begin{equation}
\label{eq:MatrixIrredRep}
D^{\{\nu\}}_{ab}(\mathscr G) ~ ,
\end{equation}
where $a, b$ are the matrix indices running from 1 to the dimension $|\nu|$ 
of the irreducible representation.

In order to compute the transition amplitude of the electric Hamiltonian,
one can use the following corollary of the Peter-Weyl theorem.
Let $\hat{C}$ denote the quadratic Casimir operator,
\begin{equation}
\label{eq:Casimir}
\hat{C} = \sum_{j=1}^{N^{2}-1} \hat{\tau}^{2}_{j} ~ .
\end{equation}
Let $C_{\nu}$ denote the eigenvalue of $\hat{C}$ in irreducible 
representation $\{\nu\}$. 
Then as a corollary of the Peter-Weyl theorem, one can express a matrix element of an operator function $f(\hat{C})$ by
\begin{eqnarray}
\label{eq:PeterWeylTheor}
&&\langle \mathscr G | f(\hat{C}) | \mathscr G' \rangle =
\nonumber \\
&&\sum_{\{\nu\}} \sum_{a,b=1}^{|\nu|} ~ \sqrt{|\nu|} ~ D^{\{\nu\}}_{ab}(\mathscr G^{-1})
~ f(C_{\nu}) ~ \sqrt{|\nu|} ~ D^{\{\nu\}}_{ba}(\mathscr G') =
\nonumber \\
&& \sum_{\nu} ~ |\nu| ~ f(C_{\nu}) ~ Tr[ D^{\{\nu\}}(\mathscr G^{-1} \mathscr G') ] =
\nonumber \\
&& \sum_{\nu} ~ |\nu| ~ f(C_{\nu}) ~ \chi^{\{\nu\}}(\mathscr G^{-1} \mathscr G') ~ .
\end{eqnarray}
The sum over $\{\nu\}$ runs over all partitions 
$\{\nu\} \equiv \{\nu_{1};\dots;\nu_{N-1}\}$ with
$\nu_{1} \ge \dots \ge \nu_{N-1}$ and 
\begin{equation}
\label{eq:Charact}
\chi^{\{\nu\}}(\mathscr G) = Tr[ D^{\{\nu\}}(\mathscr G) ]
\end{equation}
denotes the group character.

\end{document}